\documentclass[twocolumn,superscriptaddress]{revtex4-1}
\usepackage{graphicx}
\usepackage{bm}
\usepackage{amsmath}

\renewcommand{\vec}[1]{\bm{#1}}
\newcommand{\unitvec}[1]{\hat{\bm{#1}}}
\newcommand{\lbs}{\sigma}
\newcommand{\sss}{\tau}

\begin{document}
	
\author{A.~S.~Meeussen}
\affiliation{AMOLF, Science Park 104, 1098 XG Amsterdam, the Netherlands}
\affiliation{Huygens-Kamerlingh Onnes Laboratory, Universiteit Leiden, PO Box 9504, 2300 RA Leiden, The Netherlands}
\author{E.~C.~O\u{g}uz}
\affiliation{School of Mechanical Engineering and The Sackler Center for Computational Molecular and Materials Science, Tel Aviv University, Tel Aviv 69978, Israel}
\author{M.~van Hecke}
\affiliation{AMOLF, Science Park 104, 1098 XG Amsterdam, the Netherlands}
\affiliation{Huygens-Kamerlingh Onnes Laboratory, Universiteit Leiden, PO Box 9504, 2300 RA Leiden, The Netherlands}		
\author{Y.~Shokef}
\affiliation{School of Mechanical Engineering and The Sackler Center for Computational Molecular and Materials Science, Tel Aviv University, Tel Aviv 69978, Israel}

\title{Response evolution of mechanical metamaterials under architectural transformations}

\begin{abstract}
Architectural transformations play a key role in the evolution of complex systems, from design algorithms for metamaterials to flow and plasticity of disordered media. Here, we develop a general framework for the evolution of the linear mechanical response of network structures under discrete architectural transformations via sequential removal and addition of elastic elements. We focus on a class of spatially complex metamaterials, consisting of triangular building blocks. Rotations of these building blocks, corresponding to removing and adding elastic elements, introduce (topological) architectural defects. We show that the metamaterials' states of self stress play a crucial role, and that the mutually exclusive self stress states between two different network architectures span the difference in their mechanical response.	For our class of metamaterials, we identify a localized representation of these states of self stress, which allows us to capture the evolving response. We use our insights to understand the unusual stress-steering behaviour of topological defects.	 
\end{abstract}

\maketitle

\section{Introduction}\label{sec:introduction}

The unique properties of mechanical metamaterials emerge from the assembly of simple structural unit cells connected by local interactions. Targeted design of such assemblies has aided the creation of metamaterials with a broad range of responses and potential functionalities~\cite{Overvelde2016,Kang2014, Silverberg2014,Celli2018,Dudte2016c,Paulose2015,Schumacher2015,Haynes2015}.
So far, most metamaterial design has been focused on the creation of metamaterials with compatible or \textit{floppy} motions: low-energy deformations, which dominate the material's response to external probing, and lead to unusual properties such as negative Poisson ratio or vanishing shear modulus~\cite{Lakes1987,Kadic2012}. However, incompatibility or frustration offers a new avenue for designing material responses at higher energies, for example to produce materials with tunable stiffness~\cite{Coulais2018a}. Such frustration in mechanical metamaterials is closely related to other artificial frustrated systems, such as artificial spin ice~\cite{Wang2006,Nisoli2013}, colloidal ice~\cite{Libal2006,Ortizambriz2016} and colloidal antiferromagnets~\cite{Han2008,Shokef2011,Leoni2017}.\\

Recently, we presented a systematic strategy to introduce defects, and in particular topological defects, in a novel class of mechanical metamaterials~\cite{Hecke2019}. These consist of 2D triangular building blocks, and are a mechanical analogue of spin systems with tunable ferromagnetic and antiferromagnetic interactions, where the nature of the interaction is set by the orientation of the building blocks. We showed how to design a large number of compatible structures in this class---including the well-known rotating square mechanism~\cite{Grima2000,Mullin2007,Hecke2019}. We subsequently introduced (topological) defects in our metamaterials by rotating one or more building blocks. These architectural transformations affect the mechanical response and allow us to direct the stress concentration in these structures~\cite{Hecke2019}. Similarly, bond cutting strategies have recently been used to modify the elastic moduli of disordered networks~\cite{Ellenbroek2009,Ellenbroek2015,Goodrich2015}, and spatial deformations in allosteric networks~\cite{Rocks2017}. More widely, discrete changes in contact networks of flowing disordered media similarly lead to the evolution of mechanical properties~\cite{Sussman2016,Hexner2018,Hexner2018a,Bassett2015}. However, a general framework to describe the evolution of the linear response of complex spring networks under architectural transformations is lacking.\\

\begin{figure*}
	\centering
	\includegraphics[]{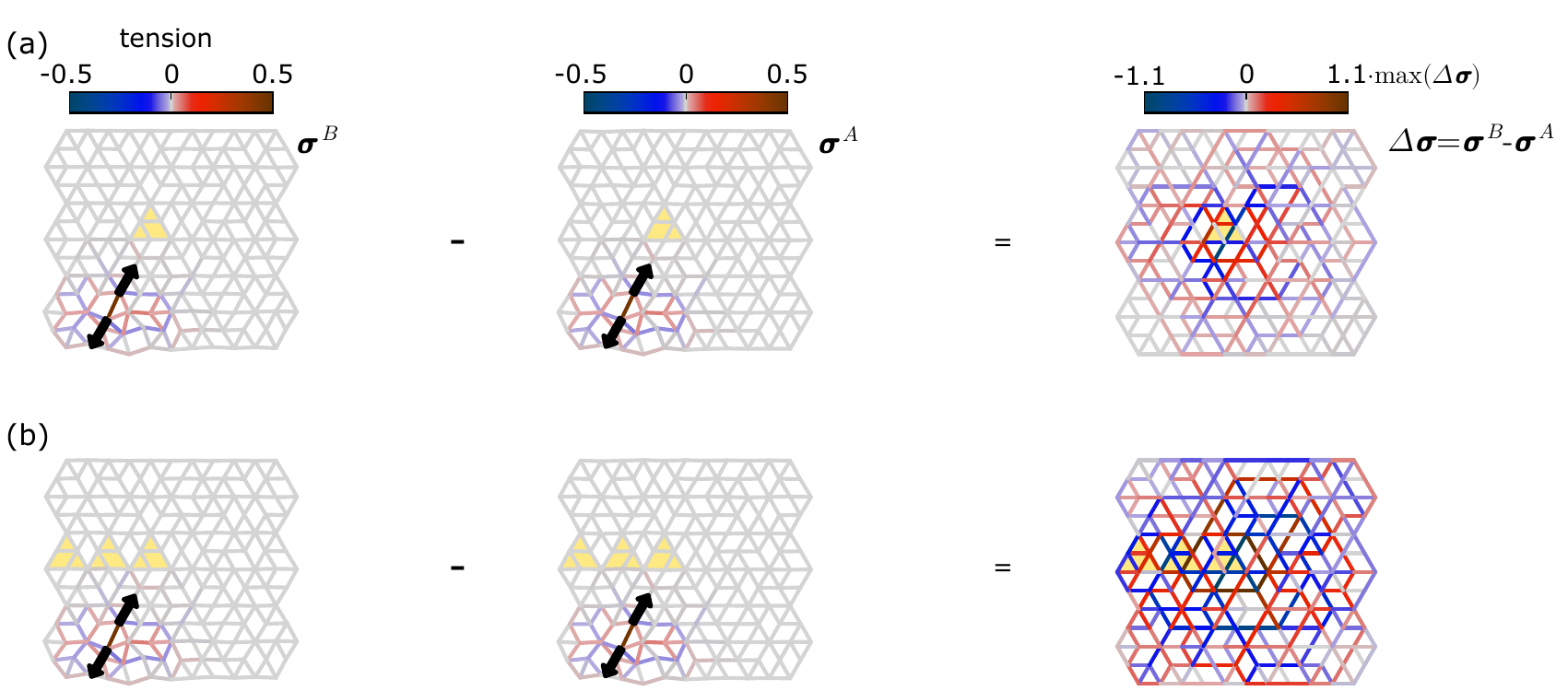}
	\caption{(a) Under the same applied load (black arrows, tripled in size for clarity), two mechanical networks differing by a small number of bonds, highlighted by the yellow triangles (left vs. centre) differ in their stress response (colour bar). Depending on the material's changing internal architecture, the stress difference ($\Delta \vec{\lbs}$, right) can be either quasilocalized when an ordinary defect is introduced (a) or diffuse if a \textit{topological} defect is created (b). The same physical principles underlie both cases: the stress difference is governed by the networks' changing states of self stress.
	}
	\label{fig_intro}
\end{figure*}

To motivate our work, consider two examples of the response evolution under architectural transformations, illustrated in Fig.~\ref{fig_intro}. The examples show two architectural transformations that produce an ordinary (Fig.~\ref{fig_intro}a) and a topological defect (Fig.~\ref{fig_intro}b) respectively. For each case, we show the stress response under an applied load before and after transformation, and focus on the stress difference as a measure of the evolution of the response. In the former case, where a single triangular building block is rotated, the stress difference is localized around the rotated block (Fig.~\ref{fig_intro}a). In the latter case, the stress difference spreads throughout the system (Fig.~\ref{fig_intro}b).\\

Our goal is to understand what controls these distinct stress differences. 
To do so, we study the linear response of spring networks under architectural transformations. The possible stress fields inside such a network form the \textit{stress space}, which is composed of load-bearing states (\textit{LB-states}), accessible via external loading, and states of self-stress (\textit{SS-states}), which are stress configurations with zero net force on all nodes. Understanding the evolution of the mechanical response entails describing the evolution of these spaces. For the overconstrained system at hand, the states of self-stress can be obtained in closed form, and we show how to use this information to completely capture the response evolution. Specifically, we find that the stress field difference between two networks as shown in Fig.~\ref{fig_intro}, is spanned by their small number of mutually exclusive SS-states. The presence of closed form SS-states in our metamaterials therefore enables us to determine a priori how small modifications in network architecture affect the mechanical response.\\

In the following sections, we discuss the linear mechanical formalism underlying our findings, which states that stress distributions inside mechanical networks under external loading are spanned by LB-states, while SS-states---which produce zero net forces---are inaccessible stress states of the network. We conclude that the stress response difference between networks with related architectures must be spanned by their mutually exclusive SS-states (Sec.~\ref{sec:linmech}). We then present our non-periodic compatible mechanical metamaterials, consisting of stacked anisotropic unit cells that can deform in harmony~\cite{Hecke2019} (Sec.~\ref{sec:metamat}), and in which the SS-space can be represented as a set of localized states (Sec.~\ref{sec:sss}). We demonstrate how sequential building-block rotations produce architectural changes that introduce controlled frustration, producing varying configurations of (topological) defects (Sec.~\ref{sec:defects}). In spite of the presence of such frustration, all SS-states can still be constructed straightforwardly (Sec.~\ref{subsec:response_evolution_sss}). As a consequence, SS-states that are not shared between any two architecturally-related networks are easily identified, and are confirmed to span the stress response difference under identical loads (Sec.~\ref{subsec:response_evolution_process_I}--\ref{subsec:response_evolution_interpretation}). Lastly, we use our knowledge of the SS-states to understand how topological defects steer stresses into different parts of a metamaterial, illustrating that our findings may be useful for designing metamaterials with targeted stress responses (Sec.~\ref{sec:design}).\\

\section{Linear mechanics: states of self stress and floppy modes}\label{sec:linmech}

In order to understand the comparative response of mechanical networks with closely related architectures, we now introduce the linear-elastic material model that underlies our findings~\cite{Pelle,Pellegrino1993}. We discuss how a mechanical metamaterial's floppy modes (FM), load-bearing stresses (LB-states), and states of self stress (SS-states) naturally arise from this theory, and show that knowledge of the SS-states suffices to understand the difference in mechanical response of two architecturally related materials.\\

We model our networks as freely hinging nodes connected by Hookean springs. The network's mechanics are described by three linear-algebraic matrix equations that relate forces exerted by each bond---which we refer to as stresses---to the net forces on and displacements of each node. First, node forces $\vec{f}$ are related to bond {stresses (or tensions), $\vec{\lbs}$ via a kinematic matrix, $\bm{R}^T$, which is constructed using the network's architectural layout, such that $\vec{f}=\bm{R}^T \vec{\lbs}$. Similarly, node displacements $\vec{u}$ map to bond elongations $\vec{e}$ via the transpose of the kinematic matrix, known as the rigidity matrix $\bm{R}$, so that $\vec{e}=\bm{R}\vec{u}$. Finally, bond elongations and bond stresses are related by a Hookean constitutive law, $\vec{\lbs} = \bm{K} \vec{e}$, where $\bm{K}$ is a diagonal matrix of spring constants, which we will set to unity in what follows. The three matrix equations above relate all possible node forces, bond stresses, bond elongations, and node displacements of the network, and thus govern the material's linear mechanical response.\\

In practice, we construct a material's kinematic matrix as follows. Consider two nodes $i,j$ in a 2D plane, connected by a bond $ij$. Their linearized elongation under planar displacements of the nodes $\vec{u}=(u_{ix}, u_{iy}, u_{jx}, u_{jy})$ is then given by $e_{ij} =[-n_{x},-n_{y}, n_{x}, n_{y}]\vec{u}$, where $\unitvec{n}$ is the unit vector along the bond running from $i$ to $j$. The $4\times1$ kinematic matrix is then given by $\bm{R}^T = [-n_{x},-n_{y}, n_{x}, n_{y}]^T$, and maps the bond's stress due to bond elongation, $s_{ij} = K e_{ij}$, to node forces $\vec{f}=(f_{ix}, f_{iy}, f_{jx}, f_{jy}) = \bm{R}^T s_{ij}$. Extending this 2D network to include $N_n$ nodes and $N_b$ bonds produces a $2N_n \times N_b$ kinematic matrix, where each of the columns corresponds to a particular bond's connection between two end nodes, as above. Therefore, the domain of the kinematic matrix is an $N_b$-dimensional space of stress vectors, in which each vector component corresponds to a bond.\\

The vector subspaces of the kinematic matrix---its kernel and row space, which form the domain, and its cokernel and column space, which form the codomain---have a particular insightful physical interpretation~\cite{Pellegrino1993}. First, the row space is spanned by the LB-states, symbolized by $\unitvec{\lbs}$, or stress eigenvectors that produce finite node forces. Secondly, if the system is overconstrained~\cite{Guest2003a}, the kinematic matrix's kernel is nontrivial and spanned by a finite number of zero eigenvectors, or bond stress configurations that lead to zero net node forces. These are the network's SS-states, symbolized by $\unitvec{\sss}$. Similarly, if the network is underconstrained, the cokernel consists of \textit{floppy modes} (FM), node displacement vectors that produce no bond elongations and thus cost no elastic energy. In two dimensions, these FM include a total of three rigid-body motions, a rotation and two translations. Lastly, the column space contains all displacement vectors that produce finite bond elongations: this column space corresponds one-to-one to the LB-states of the row space. Thus, the SS-space and LB-space together span the entire space of possible bond stress configurations---the former being inaccessible states, and the latter supported states---and they therefore govern the network's response to external loading.\\

While the subspaces' bases are often not simple to determine, their dimensions follow directly from the rank-nullity theorem that relates the subspace dimensions of the network's kinematic matrix~\cite{Maxwell1864,Calladine1978a,Connelly1982RigidityAE,Pelle,Lubensky2015}. The rank-nullity theorem states that the sum of the number of independent FM ($N_{FM}$) and the number of independent LB-states is equal to $2N_n$, while the sum of the number of independent SS-states ($N_{SSS}$) and LB-states must be equal to $N_b$. Therefore, the difference between the number of SS-states and FM has a consistent expression for all 2D materials: 
\begin{equation}\label{eq:index}
	\nu = N_{FM} - N_{SSS} -3= 2N_n - N_b-3,
\end{equation}
where the final term of $-3$ represents the three trivial rigid-body motions in 2D, so that $N_{FM}$ includes only internal floppy deformations of the structure.\\

The above linear-elastic model helps understand the difference in stress response between two networks with closely related architectures that differ by a small number of bonds, but have the same number and spatial configuration of nodes. In either network, the SS-space and LB-space together span the entire space of possible bond stress configurations. Some SS-states and LB-states are shared between the two materials, while others are unique to either of the pair. Any SS-state unique to one network must be an LB-state---up to stresses on the networks' distinct bonds---in the other structure. Since the stress response of any network is a linear combination of its LB-states, the stress response difference between the two networks must therefore lie in the space spanned by their unique, non-shared SS-states. In other words, \textit{with knowledge of the mutually exclusive SS-states of two mechanical networks, we can a priori determine how their stress response differs under arbitrary external loading.}\\

We note here that our analysis concerns the material's response under an applied \textit{supported} load: external forces that actuate a floppy motion of the material lead to an indeterminate response~\cite{Pelle}, which we do not consider here.\\

\section{Structurally complex mechanical metamaterials}\label{sec:metamat}
We now demonstrate the efficacy of predicting the stress response difference using SS-states---an approach valid for \textit{any} mechanical network architecture---in a particular class of structurally complex mechanical metamaterials~\cite{Hecke2019}. Their specific architecture allows us to easily enumerate and construct a basis of SS-space consisting of highly spatially localized states, and we show later that this complete description of SS-space produces a direct prediction of the stress response difference between two networks of differing designs under identical, external, supported loads.\\

\begin{figure*}
	\centering
	\includegraphics[]{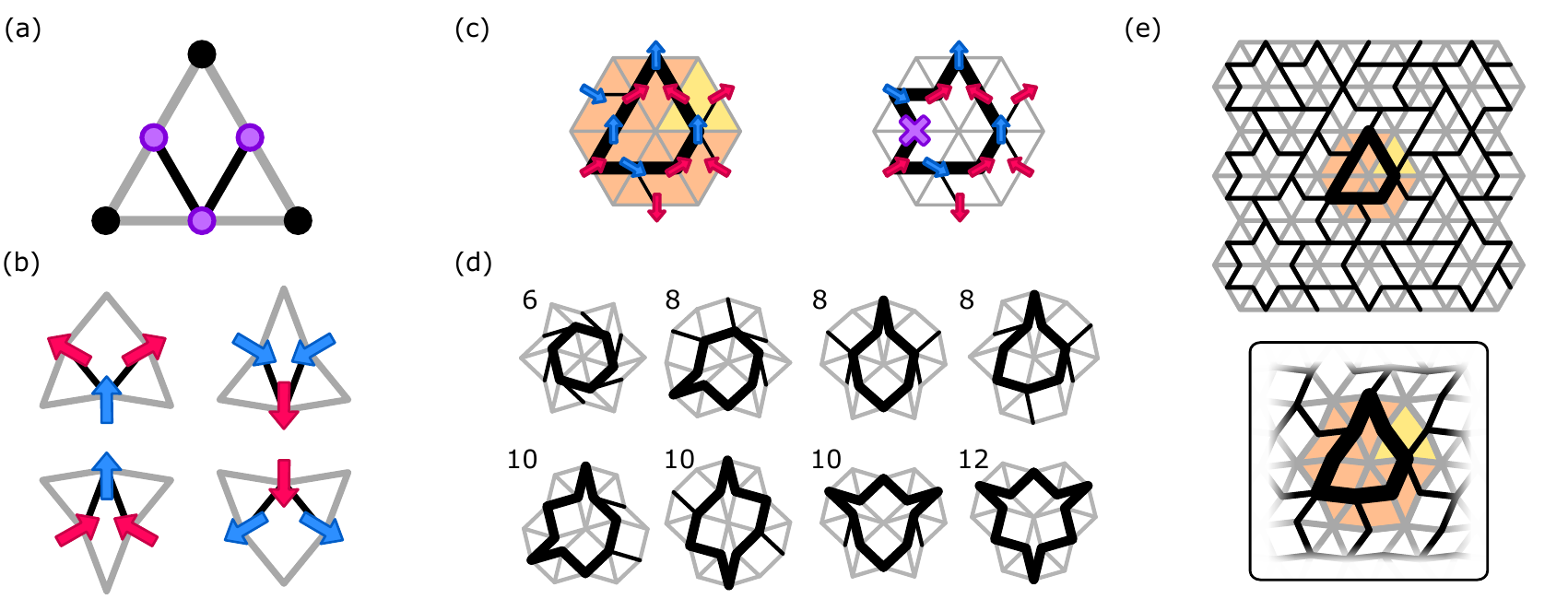}
	\caption{
		(a) Our mechanical building block, or supertriangle, consists of three corner nodes (black circles) and three internal nodes (purple), connected by a perimeter of edge bonds (grey lines). The internal nodes are connected by two internal bonds (black).
		(b) Internal node motions and internal bonds map to Ising spins and antiferromagnetic interactions. Compatible deformations of the supertriangle correspond to ground states of this Ising model. The correspondence between spin states (`in' and `out', indicated with blue and red arrows) and motion of the internal nodes is opposite for upward- and downward-pointing building blocks.
		(c) Supertriangles (yellow triangle) are stacked together to create a superhexagon. Superhexagons contain a closed \textit{local loop} of internal bonds (bold black lines). The metamaterial deforms harmoniously only when a spin ground states exists that satisfies all antiferromagnetic constraints simultaneously. Local loops with an even number of bonds satisfy this requirement (left): the superhexagon is compatible. Incompatible superhexagons have  loops containing an odd number of bonds (right) that frustrate at least one antiferromagnetic interaction (purple cross). The odd local loop represents a \textit{defect} in the system.
		(d) The eight possible even local loop shapes (number of bonds indicated) in a compatible superhexagon are shown (bold black lines). Internal bonds outside the local loop were chosen arbitrarily, and the corresponding floppy modes are illustrated as deformations of the superhexagons.
		(e) A large compatible metamaterial is created by stacking building blocks, ensuring that the local loops inside each superhexagon (orange hexagon) contain an even number of bonds. The compatible metamaterial deforms harmoniously (zoom-in).}
	\label{fig_metamat}
\end{figure*}

Our complex mechanical metamaterials are assembled by stacking together copies of an anisotropic triangular building block~\cite{Hecke2019} (Fig.~\ref{fig_metamat}a) that we will refer to as a \textit{supertriangle}. The supertriangle consists of six Hookean \textit{edge bonds}, connected in a triangular shape. Three freely pivoting \textit{corner nodes} connect the bonds at the triangle's corners, while three \textit{internal nodes} connect the sides. The supertriangle is made anisotropic by connecting two of the internal node pairs with two additional Hookean \textit{internal bonds}, leaving the third pair unconnected. This building block exhibits a local FM: a \textit{compatible} internal deformation that does not deform any of the rigid bonds (Fig.~\ref{fig_metamat}b).\\

The smallest nontrivial structure, made with six supertriangles, is a hexagonal stack or \textit{superhexagon} (Fig.~\ref{fig_metamat}c). Such stacks are called \textit{compatible} when there is a collective FM, such that all individual supertriangles can deform according to their local FM simultaneously; otherwise, the stack is \textit{incompatible} or frustrated. Evidently, even though the number of nodes and bonds of compatible and incompatible superhexagons are identical ($N_n=19$ and $N_b=8$), they show distinct mechanical behaviour. Using Eq.~(\ref{eq:index}), we find that incompatible superhexagons have no FM and a single SS-state, while compatible superhexagons have a single FM and two SS-states.\\

To obtain clear design rules for compatibility, we map the local FM of a supertriangle to the ground state of an Ising model with antiferromagnetic interactions~\cite{Hecke2019}. Specifically, each internal node corresponds to a spin site, while each internal bond represents an antiferromagnetic interaction. Spins may be in an `out' state or an `in' state; mechanically, this corresponds to an outward or inward motion of the internal nodes with respect to the centre for upward-pointing supertriangles (and vice versa for downward-pointing supertriangles) indicated by the red and blue arrows in Fig.~\ref{fig_metamat}b. The supertriangle's mechanical FM then corresponds uniquely to a spin configuration that satisfies both antiferromagnetic interactions: the internal bonds connect spin sites at two internal nodes in opposite states, while nodes not connected by an internal bond both move inward (or both outward), representing two ferromagnetically interacting spins.\\

For a compatible superhexagon, the spin orientations of all adjacent supertriangles have to match up exactly. Figure~\ref{fig_metamat}c demonstrates that the internal bonds inside a superhexagon form a closed \textit{local loop} corresponding to a ring of antiferromagnetic interactions. The supertriangles collectively deform harmoniously and compatibly if, and only if, the corresponding antiferromagnetic Ising model is in a ground state, so that each antiferromagnetic interaction connects two spins in opposite states. This requirement is only met if the local loop contains an even number of interactions. Hence, a superhexagon is only compatible if the local loop contains an even number of internal bonds (Fig.~\ref{fig_metamat}c, left).\\

By contrast, when the local loop has an odd number of internal bonds, the superhexagon is geometrically frustrated and incompatible~\cite{Wannier1950,Toulouse1977}. In the Ising model language, there is then always an antiferromagnetic interaction that cannot be satisfied (Fig.~\ref{fig_metamat}c, right), so that the odd local loop represents a defect in the mechanical system.\\

We note here that this mapping to an Ising model with binary states is complete only for compatible metamaterials which posses a FM in which displacements alternate in direction and all have the same magnitude. As we will show below, in incompatible situations, the magnitude of the displacements varies continuously with position and then this mapping to the Ising model serves only to demonstrate whether or not there exists a compatible deformation.\\

In Fig.~\ref{fig_metamat}d, we show the FM in compatible superhexagons for each of the eight possible even local loop shapes (with six, eight, ten or twelve bonds, bold black lines); the FM is present independently of the choice of internal bonds outside the local loop (thin black lines).\\

Metamaterials consisting of large stacks containing many supertriangles (Fig.~\ref{fig_metamat}e) typically contain many superhexagons, each sporting a local loop of internal bonds. Designing the material so that there are only even local loops in the system ensures that all superhexagons are compatible, the material has a single global FM, and can deform harmoniously. Conversely, odd local loops generate geometric frustration and incompatibility, resulting in the absence of a global FM. As shown in previous work~\cite{Hecke2019}, there is an extensive number of metamaterial designs made of these supertriangular building blocks. Moreover, we can design a wide array of geometries with varying isotropy, auxeticity, and periodicity. Here, we explore the evolving mechanical response under architectural changes in this class of spatially complex metamaterials, and our findings thus hold for metamaterials with a wide range of mechanical properties.
	
\section{States of self stress in superhexagons and larger metamaterials}\label{sec:sss}

We now show how to identify the dimension and shape of the SS-space in our complex metamaterials, which governs the differential response of architecturally related networks. Our compatible metamaterials have one global FM by construction, while frustrated ones have none. Hence, to obtain the number of independent SS-states from Eq.~(\ref{eq:index}), it suffices to calculate the index $\nu$. We show below that $\nu$ follows directly from the number $H$ of superhexagons contained inside our metamaterial, and that each compatible (incompatible) superhexagon contains two (one) localized SS-states that can be explicitly and straightforwardly constructed.\\

\begin{figure}
	\centering
	\includegraphics[]{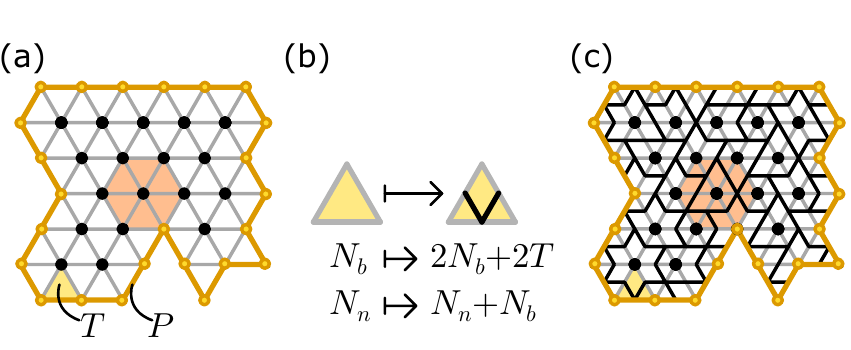}
	\caption{
		The number of nodes and bonds in a metamaterial can be counted exactly. (a) Starting from a network of T adjacent triangular cells (yellow triangle) with a perimeter of $P$ bonds and $P$ nodes (orange lines and circles), the number of nodes and bonds $N_n$ and $N_b$ can be counted exactly. Each internal lattice point (black circles) is surrounded by a hexagon of six triangular blocks (orange hexagon).
		(b) Each block is decorated with two internal bonds and three internal nodes, producing a supertriangle.
		(c) This decoration produces a metamaterial. The number of nodes and bonds increases to $N_n+N_b$ and $2N_b+2T$.
	}
	\label{fig_counting}
\end{figure}

To count the number of superhexagons in a metamaterial, we first focus on the structure's scaffold that consists of corner nodes connected by a triangular lattice (Fig.~\ref{fig_counting}a). If such a scaffold contains $T$ triangles and a perimeter of $P$ bonds, it contains
\begin{equation}\label{eq:hexcount}
H = \frac{T-P}{2}+1
\end{equation}
full hexagons of six triangles, each surrounding a distinct bulk corner node (orange hexagon and bold black dots in Fig.~\ref{fig_counting}a). This expression is derived as follows: a single triangle has $T=1$, a perimeter of $P=3$ and $H=0$ hexagons. Adding a triangle to an existing system increases the number of triangles by one ($T \to T+1$), and either increases the perimeter by two bonds and produces no new hexagon ($P \to P+2, H \to H$), or increases the perimeter by one bond and produces a new hexagon ($P \to P+1, H \to H+1$). By induction, Eq.~(\ref{eq:hexcount}) then holds for all lattices.\\

\begin{figure*}
	\centering
	\includegraphics[]{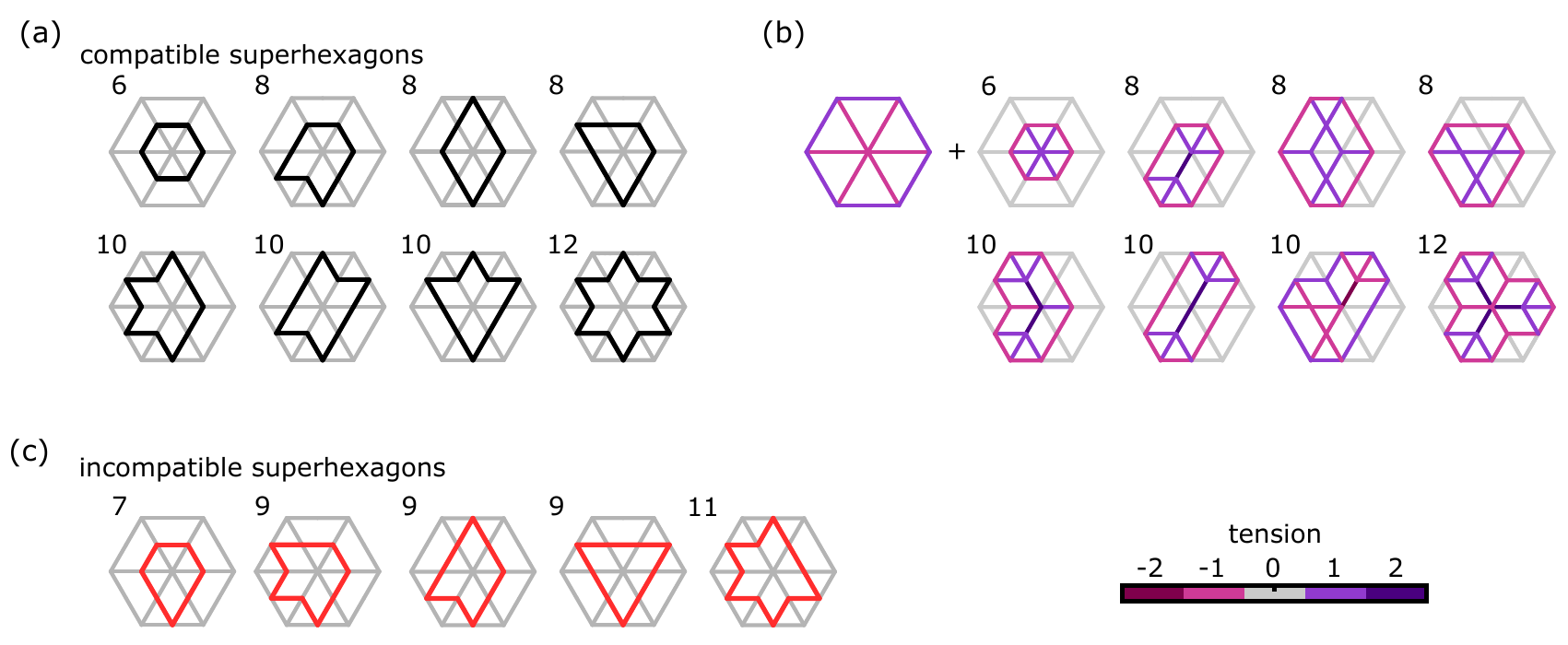}
	\caption{
		States of self stress (SS-states) are localized in superhexagons.
		(a) The eight possible even local loop shapes in a compatible superhexagon are shown (black lines).  The number of bonds in each loop is indicated. Some internal bonds outside the local loop may be chosen freely (not shown here for clarity), while the triangular scaffold (grey solid lines) is always present.
		(b) The compatible superhexagons contain two non-orthonormal SS-states (colours): a \textit{radial SS-state} localized on the triangular scaffold (left), and a \textit{loop SS-state} fully localized on internal bonds in the local loop and the triangular scaffold.
		(c) Incompatible superhexagons contain odd local loops that come in five distinct shapes (red lines). These structures each support only the radial SS-state.
	}
	\label{fig_localsss}
\end{figure*}

We now use this information to determine a general expression for $\nu$ in our metamaterials. Adding two internal bonds and three internal nodes to every triangle in the scaffold---thus creating a stack of $T$ supertriangles---generates a metamaterial (Fig.~\ref{fig_counting}b). Since the triangular scaffold contains a total of $N_b=\frac{3T+P}{2}$ bonds, the metamaterial will contain $3T+P$ edge bonds and an additional two internal bonds per triangle, yielding a total of $N_b=5T+P$ edge and internal bonds. In addition, the scaffold contains $N_n=\frac{T+P}{2}+1$ corner nodes; the metamaterial has an additional three internal nodes that are shared between two triangles, unless they lie on the structure's perimeter. This yields a total of $N_n=2T+P+1$ corner and internal nodes in the metamaterial (Fig.~\ref{fig_counting}c). The metamaterial's index $\nu$ is thus equal to
\begin{equation}
\nu = 1- 2H.
\end{equation}\\

From Eq.~(\ref{eq:index}), and using the fact that the number of FM in a metamaterial is either one or zero, we obtain an exact expression for the dimension of SS-space in our metamaterials: $N_{SSS} = 2H$ in compatible systems, and $N_{SSS} = 2H-1$ in incompatible ones. This expression is consistent with our finding in Sec.~\ref{sec:metamat} that a compatible superhexagon contains two SS-states, while an incompatible superhexagon has one SS-state. Thus, in a compatible metamaterial with $H$ hexagons, we can identify $2H$ independent SS-states localized within each of the metamaterial's superhexagons; these SS-states exactly span the $2H$-dimensional SS-space. Therefore, \textit{all independent SS-states of a compatible metamaterial can be constructed as localized states within each of the larger metamaterial's superhexagons}.\\

We illustrate the compact, superhexagon-localized representation of all independent SS-states in Fig.~\ref{fig_localsss}. Consider a metamaterial consisting of a single, compatible superhexagon. Its local loop contains an even number of internal bonds; the structure has a single FM, and two SS-states. Figure~\ref{fig_localsss}a enumerates the eight possible even local loop shapes (up to rotations and reflections); internal bonds outside of the local loop do not carry stress in any of the SS-states, and are not shown for clarity. Due to the network's highly regular geometry, the SS-states are found by inspection to have a simple structure: one \textit{radial SS-state} is independent of the superhexagon's internal bonds and is purely supported on edge bonds, while the other \textit{loop SS-state} involves  the internal bonds of the local loop (Fig.~\ref{fig_localsss}b). The location of internal bonds that are not part of the local loop are irrelevant for both the radial and loop SS-states. Bond stresses of both radial and loop SS-states are integer multiples due to the underlying building blocks' six-fold rotational symmetry. By contrast, a single, incompatible superhexagon containing an odd local loop has no FM and only one SS-state; the local loop has five possible shapes (Fig.~\ref{fig_localsss}c), and the superhexagon supports only the single radial SS-state (Fig.~\ref{fig_localsss}b, left).\\

In compatible metamaterials consisting of $H$ compatible superhexagons, the $2H$-dimensional SS-space is therefore spanned by $H$ radial and $H$ loop SS-states, each of which is localized to a single superhexagon.  Similarly, in a metamaterial with a single incompatible superhexagon, the $2H-1$-dimensional SS-space consists of the $H$ radial SS-states, and the $H-1$ loop SS-states in the remaining compatible superhexagons. For larger numbers $H_o >1$ of incompatible superhexagons, $H$ radial and $H-H_o$ loop SS-states are present in the network, with the remaining $H_o-1$ SS-states not localized to a single superhexagon. \\

\section{Architectural defects}\label{sec:defects}
While we can make a large variety of compatible metamaterials (a number that grows exponentially with the number of supertriangles in the structure)~\cite{Hecke2019}, an even larger amount of frustrated designs exist that cannot deform harmoniously due to the presence of one or more odd local loops. The mechanical frustration induced by such defects generally produces undesired effects when their presence is not controlled, such as decay of a desired FM~\cite{Filipov2015,Coulais2018a}, or structural failure when frustration-induced bond stresses exceed the bond buckling threshold~\cite{Gaitanaros2012}. However, when frustration is introduced in a controlled and well-understood manner, it may be harnessed to design desirable or unusual physical properties, such as localized buckling zones~\cite{Paulose2015,Kang2014,Hecke2019}, or geometric frustration in spin-ices~\cite{Toulouse1977,Ortizambriz2016,Nisoli2017}.\\

\begin{figure*}
	\centering
	\includegraphics[]{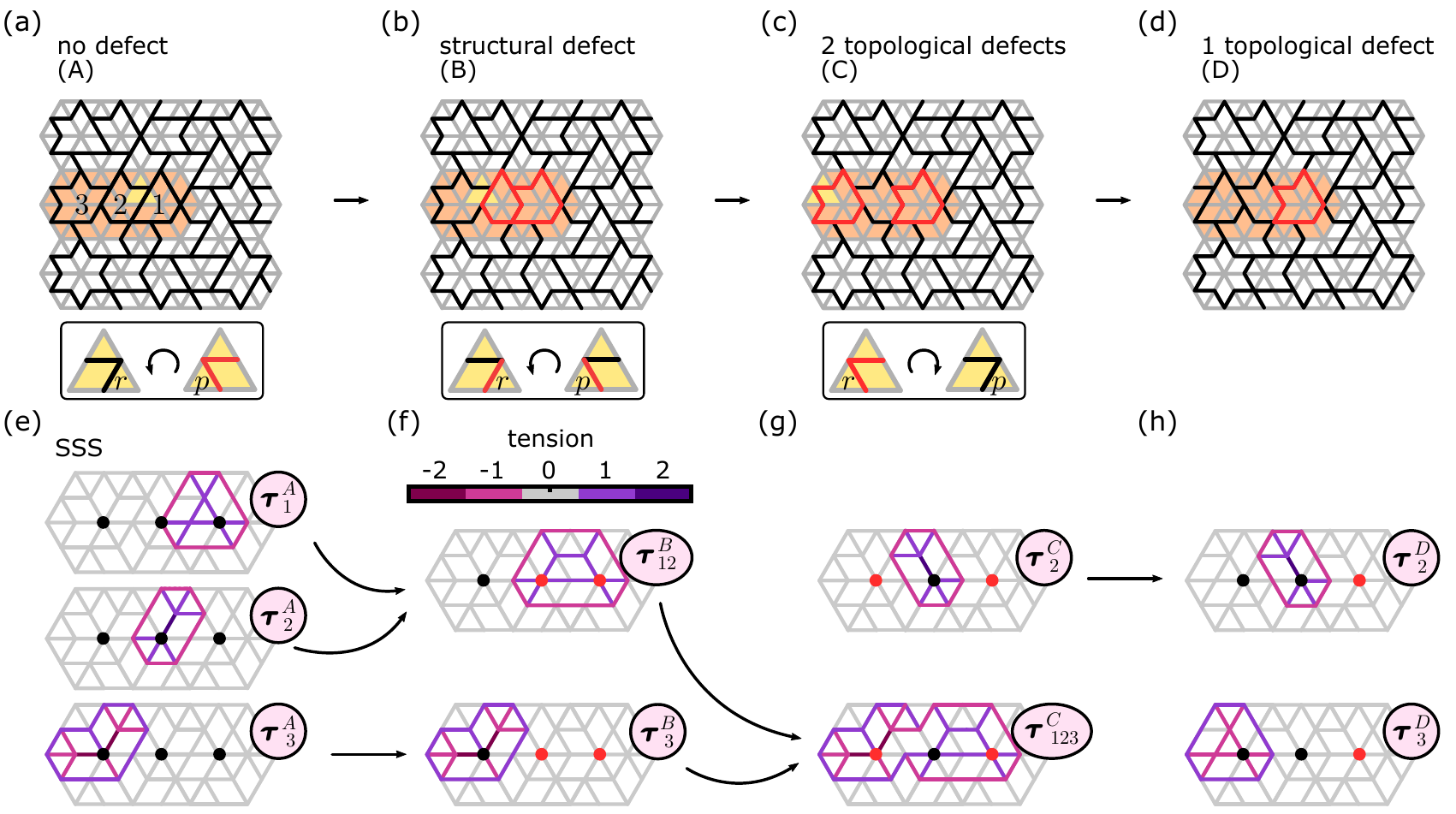}
	\caption{
		(a) A compatible metamaterial (no defect, A) contains only even local loops (internal bonds highlighted in black). Three superhexagons ($1,2,3$; orange) that change parity under consecutive transformations are highlighted. Rotating a single bulk supertriangle shared by superhexagons $1$ and $2$ (yellow triangle, inset) removes a bond $r$ and adds a bond $p$. 
		(b)  The supertriangle rotation generates two adjacent odd local loops (red lines). These form a \textit{structural defect} (B) that frustrates the compatible motion of the material. The adjacent odd local loops are moved apart by selecting and rotating a second supertriangle in superhexagons $2$ and $3$  (inset). 
		(c) Two \textit{topological defects} (C), or isolated odd local loops, are created: an even local loop now separates the odd local loops. A final rotation in superhexagon $3$ (inset) removes one of the odd local loops from the material. 
		(d) A single topological defect (D) remains.
		(e) The three numbered superhexagons in the compatible metamaterial are shown, along with their central corner nodes (black circles) and their corresponding loop SS-states $\vec{\sss}_1^A$, $\vec{\sss}_2^A$ and $\vec{\sss}_3^A$ (colours).
		(f-h) Transforming the network to produce a structural defect, two topological defects, and a single topological defect (central corner nodes of incompatible superhexagons indicated in red) results in a sequential evolution where new SS-states are formed from linear combinations of old SS-states (arrows; see text for detailed expressions). In panels (f) and (g), two odd local loops are present in the network, and the SS-space can no longer be represented by purely superhexagon-localized SS-states. However, a (maximally) localized representation does exist, where an SS-state runs over the superhexagons along the shortest path between the two odd local loops. In panel (h), there is only one incompatible superhexagon; all SS-states are localized within distinct superhexagons.
	}
	\label{fig_defects}
\end{figure*}

We now show how to control the frustration in our mechanical metamaterials by rotating select supertriangles in an initially compatible network. Figure~\ref{fig_defects}a shows a compatible structure with no defects (\textit{A}), where all superhexagons have even local loops (black lines). Selecting and rotating a particular supertriangle in the material's bulk (Fig.~\ref{fig_defects}a, inset) effectively removes one of the supertriangle's internal bonds---bond $r$---from the network and replaces it with a newly added internal bond $p$. The bond $r$ is part of exactly two local loops. In general, exchanging bond $r$ for bond $p$ changes the parity of these two local loops. Here, since we start from a compatible structure, rotating a supertriangle creates two adjacent odd local loops (Fig.~\ref{fig_defects}b). We will refer to such a pair of adjacent odd local loops as a \textit{structural defect} (network \textit{B}), since the odd loops may be removed by locally rotating a single supertriangle~\cite{Hecke2019}.\\

Metamaterials containing a single incompatible superhexagon can also be constructed, and have been shown to have a topological signature~\cite{Hecke2019}. Such \textit{topological defects} (network \textit{C}) can be generated from an initially compatible system via a sequence of supertriangle rotations running in a chain between the defect locus and the system's boundary. Specifically, we rotate a supertriangle at the edge of a structural defect, ensuring that this supertriangle contributes an internal bond to one odd and one even local loop (Fig.~\ref{fig_defects}b). As before, the rotation changes the parity of the two local loops it contributes to. Consequently, the two odd local loops are no longer adjacent after the transformation: they are now separated by a single even local loop. This defect configuration, consisting of two incompatible superhexagons separated by one or more compatible ones, is a complex of two \textit{topological defects} (network \textit{C}): the odd local loops can no longer be removed by a single, local supertriangle rotation. To finally obtain a single topological defect, we repeat the above procedure to displace one of the odd local loops closer and closer to the system's boundary. Finally, we select a boundary supertriangle that contributes to exactly one odd local loop, so that its rotation causes the odd loop's parity to become even  (Fig.~\ref{fig_defects}c). This transformation leaves us with an isolated incompatible superhexagon in the system's bulk, that can only be removed by an extensive number of supertriangle rotations, and that we therefore refer to as a topological defect (Fig.~\ref{fig_defects}d).\\

Supertriangle rotations thus form the minimal architectural transformations that allow us to convert one metamaterial design to any other. By a series of sequential supertriangle rotations, we can thus obtain metamaterial architectures with any desired number of frustrated odd local loops, starting from a compatible structure containing only even local loops.\\

\section{Response evolution under architectural transformations}\label{sec:response_evolution}

Starting from an initially compatible metamaterial, supertriangle rotations form minimal architectural transformations that generate predictable defect configurations. Here, we investigate how the concomitant frustration manifests in the mechanical response. Clearly, a frustrated metamaterial cannot deform harmoniously, so external forcing will generate stresses and elastic deformations. We want to understand where these stresses are localized, and how they relate to the sequence of architectural transformations that generate a given network design.\\
	
In Sec.~\ref{sec:linmech}, we discussed how the mechanical response of a network is determined by its $N_b$-dimensional stress space, which can be decomposed into two mutually orthogonal sub-spaces: the $N_{SS}$-dimensional SS-space, and the $N_{LB}$-dimensional LB-space. To understand how architectural changes affect the stress response, we therefore need to establish how the SS-space and the complementary LB-space change under architectural modifications. Our metamaterials, with their readily constructed SS-states, are especially suitable to address such general questions.\\

To capture the changes of the SS- and LB-spaces due to architectural modifications, we repeatedly use a number of basic principles that we outline here. We only consider architectural changes that consist of sequences of supertriangle rotations, and break up each supertriangle rotation into a step-by-step process where we first remove a bond and then add a bond at a different location, which simplifies our calculations and generalizes easily to other network architectures.\\

Supertriangle rotations can mutate the compatibility of our metamaterials: there exist three different mutation processes. First of all, in process I, a compatible system $A$ transforms into an incompatible system $B$ (see e.g. Fig.~\ref{fig_defects}a,b). Secondly, process II converts an incompatible system $B$ into a distinct incompatible system $C$ (see e.g. Fig.~\ref{fig_defects}b,c), and lastly, process III converts a compatible system $A$ into a compatible system $A'$. Process III can only occur for specific supertriangle rotations at the edge of a metamaterial, and is trivial from the perspective of the mechanical response; we do not consider it further here (see Appendix~\ref{app:stress_space_evolution} for details). In process I, we start from a compatible system $A$, then remove a bond labelled $r$ to obtain the intermediate system $AB$, and then add bond labelled $p$ to obtain the incompatible system $B$. In process II, we start from an incompatible system $B$, then remove a bond labelled $r$ to obtain the intermediate system $BC$, and then add bond labelled $p$ to obtain the incompatible system $C$.\\

Now that we have broken down possible structural changes into a precise sequence of removing and adding bonds, we can determine how the \textit{dimension} of the SS- and LB-space changes in each transformation step, using constraint counting (see Sec.~\ref{sec:sss}). First of all, in process I, step $A\rightarrow AB$ removes one SS-state, while  the number of LB-states remains constant. Step $AB\rightarrow B$ leaves the SS-states unaffected, while the number of LB-states increases by one. Secondly, in process II, step $B\rightarrow BC$ removes one SS-state, while  the number of LB-states remains constant. Step $BC\rightarrow C$ adds one SS-state, while the number of LB-states remains constant.\\

\begin{figure*}
	\centering
	\includegraphics[]{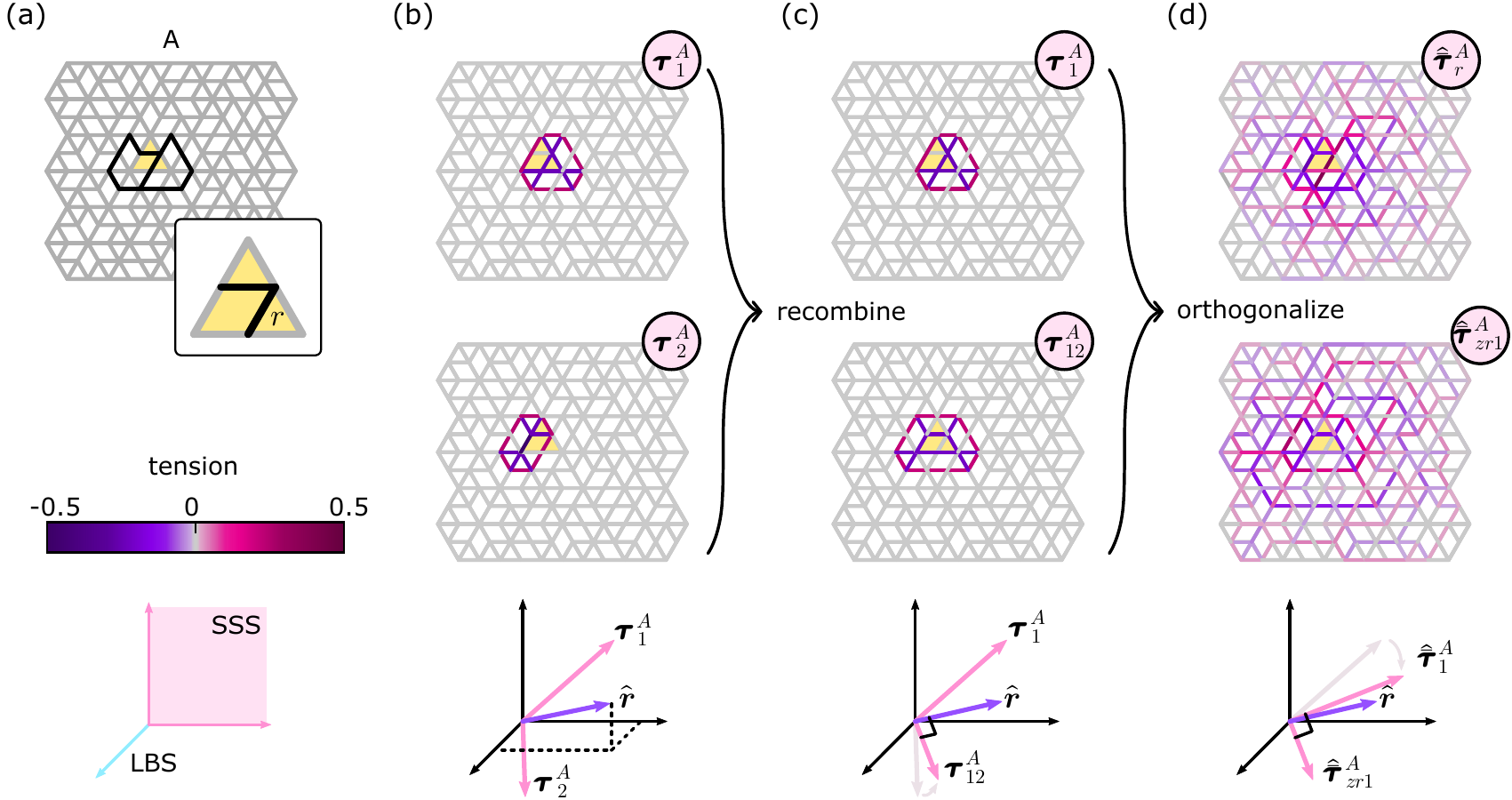}
	\caption{
		We identify the unique state of self stress (SS-state) $\unitvec{\bar{\sss}}_r^A$ that is modified under a supertriangle rotation in an initially compatible network $A$.
		(a) We transform the network by rotating a certain building block (yellow triangle), such that bond $r$ (zoom-in) is removed from the network. Bottom: the $N_b$-dimensional space of bond stress states is schematically represented as a space consisting of LB-states (blue, represented as a one-dimensional line) and SS-states (pink).
		(b) Only the two loop SS-states $\vec{\sss}_1^{A}$ and $\vec{\sss}_2^{A}$ with a nonzero stress on bond $r$ need to be considered. Bottom: the non-orthogonal SS-states $\vec{\sss}_1^{A}$ and $\vec{\sss}_2^{A}$ lie in the SS-space plane (pink vectors), while the stress vector $\unitvec{r}$ (purple vector), with nonzero stress on bond $r$, overlaps with both SS-space and LB-space (dashed lines). Both SS-states overlap with $\unitvec{r}$.
		(c) The two SS-states are recombined to yield the vectors $\vec{\sss}_1^{A}$ and $\vec{\sss}_{12}^{A}$, so that $\vec{\sss}_1^{A}$ is the only SS-state with nonzero stress on bond $r^A$. Bottom: the SS-states are recombined so that $\vec{\sss}_{12}^{A}$ is orthogonal to $\unitvec{r}$, and only $\vec{\sss}_1^{A}$ overlaps with $\unitvec{r}$.
		(d) The two SS-states are orthogonalized with respect to all other (superhexagon-localized) SS-states via a Gram-Schmidt process. Two SS-states $\unitvec{\bar{\sss}}_{r}^A$ and $\unitvec{\bar{\sss}}_{zr1}^A$ are obtained, such that only the former has nonzero stress on bond $r$. Thus, $\unitvec{\bar{\sss}}_r^A$ is lost after the supertriangle rotation that removes bond $r$. Bottom: orthogonalization produces the SS-state $\unitvec{\bar{\sss}}_{r}^A$, orthogonal to all LB-states and the remaining SS-states, and uniquely overlapping with $\unitvec{r}$.
	}
	\label{fig_sssortho}
\end{figure*}	

Crucially, changes to the \textit{dimensionality} of the SS- and LB-spaces do not capture their full reconfiguration. As an example, consider step $A\rightarrow AB$, where bond $r$ is removed from network $A$: while the number of LB-states remains constant, the removal of bond $r$ induces changes to the structure of these states. After all, LB-states may have a finite stress on bond $r$ in network $A$, but LB-states of network $AB$ must have zero stress on the nonexistent bond $r$.\\	

In order to fully capture changes in the SS- and LB-spaces, we must construct appropriate bases for them, to make their evolution tractable. As the SS-states are easier to identify than the LB-states in our particular metamaterials, we construct an orthonormal basis for the SS-space of our metamaterials, such that removing a bond $b$ will affect at most one basis vector. This basis consists of \textit{(i)} at most one SS-state vector that has a finite stress on bond $b$, which is modified under removal of bond $b$, and \textit{(ii)}
all other basis vectors that have zero stress on bond $b$~\cite{Lerner2018}.\\

The two subspaces \textit{(i)-(ii)} are mutually orthogonal; moreover, the LB-space is orthogonal and complementary to the SS-space. Hence, changes in the subspace $\textit{(i)}$ directly affect the LB-space. The LB-space ultimately determines the metamaterial's response under external loading. However, as we discussed at the end of Sec.~\ref{sec:linmech}, the stress response difference between two networks related by a single supertriangle rotation is determined by their mutually exclusive SS-states. Thus, the evolution of the SS-space suffices to capture the evolution of the metamaterial's response, as a detailed derivation in Appendices~\ref{app:stress_space_evolution}--\ref{app:stress_response_difference} confirms.\\

In the following, we therefore first describe how to construct all SS-states in compatible and incompatible metamaterials as linear combinations of radial and loop SS-states in Sec.~\ref{subsec:response_evolution_sss}. We consider process I in Sec.~\ref{subsec:response_evolution_process_I}, identifying the changes to the SS-space, and process II in Sec.~\ref{subsec:response_evolution_process_II}, again determining changes to the SS-space. Ultimately, we establish that the evolution of SS-space under supertriangle rotations is limited to a small and predictable span of stress vectors. We close this section with a discussion in Sec.~\ref{subsec:response_evolution_interpretation} of the mechanical consequences of these SS-space changes due to supertriangle rotations.

	\subsection{Constructing the states of self stress}\label{subsec:response_evolution_sss}
	
	As shown in Sec.~\ref{sec:sss}, the SS-space of any compatible metamaterial is spanned by superhexagon-localized radial and loop SS-states (see Fig.~\ref{fig_localsss}b). Together, the superhexagon-localized states form a complete, non-orthogonal basis of the material's SS-space. However, a different approach is needed to identify a complete basis of the SS-space for incompatible metamaterials: as we will show below, in frustrated systems, some SS-states cannot be represented as superhexagon-localized states, but must be \textit{delocalized}. Here, we present an iterative approach to construct a basis of SS-space for \textit{any} metamaterial---compatible or not---and show that all delocalized SS-states can be constructed as linear combinations of radial and loop SS-states.\\	

	We illustrate our approach by constructing a basis of the SS-space in the four architecturally related networks presented in Fig.~\ref{fig_defects}a-d, with network $A$ containing no defect, $B$ a structural defect, $C$ two topological defects, and $D$ a single topological defect, as a specific demonstration of our general strategy. Figure~\ref{fig_defects}e shows the three highlighted compatible superhexagons, numbered $1$, $2$ and $3$ in the compatible network $A$, that are modified during the network transformations. The three superhexagons support three radial SS-states (see Fig.~\ref{fig_localsss}b above), not shown here for brevity. As the network transformations considered here leave the scaffold of edge bonds intact, the $H$ radial SS-states remain, irrespective of the number of supertriangle rotations.
	We focus on the loop SS-states that are localized in these three superhexagons, which we will denote $\vec{\sss}^{A}_1$, $\vec{\sss}^{A}_2$, and $\vec{\sss}^{A}_3$, and which are shown in Figure~\ref{fig_defects}e. Rotating a supertriangle in network $A$ that is part of both superhexagons $1$ and $2$ removes one bond, $r$ (Fig.~\ref{fig_defects}a,b). This rotation also lowers the number of SS-states by one. First, we note that  $\vec{\sss}_3^A$ does not induce a stress on bond $r$, so that this SS-state is retained in network $B$. However, $\vec{\sss}_1^A$ and $\vec{\sss}_2^A$ do include a stress on bond $r$: hence, they cannot be SS-states of network $B$. We construct a new SS-state for network $B$ as a linear combination of $\vec{\sss}_1^A$ and $\vec{\sss}_2^A$ that leaves bond $r$ unstressed: $\vec{\sss}^{B}_{12} = \vec{\sss}^{A}_1 + \vec{\sss}^{A}_2$ (see Fig.~\ref{fig_defects}f)). Here we use the subscript $12$ to indicate that this SS-state is delocalized: it is contained within the two incompatible superhexagons 1 and 2. All other SS-states in network $A$, similar to  $\vec{\sss}^{B}_3=\vec{\sss}^{A}_3$, are retained in network $B$.\\
	
	A second supertriangle rotation in network $B$ produces two separated topological defects in network $C$ (Fig.~\ref{fig_defects}c), but does not change the number of SS-states. Since a distinct bond $r$ is now removed during the supertriangle rotation, and both $\vec{\sss}^{B}_{12}$ and $\vec{\sss}^{B}_3$ produce a finite stress on bond $r$, these two SS-states cannot persist in the network. By a similar superposition as above, we obtain a new SS-state $\vec{\sss}^{C}_{123} =  \vec{\sss}^{B}_{12} + \vec{\sss}^{B}_3$. This SS-state spans the connecting path between the two odd loops, since $\vec{\sss}^{C}_{123} =  \vec{\sss}^{A}_1 + \vec{\sss}^{A}_2 + \vec{\sss}^{B}_3$. However, to maintain the overall number of SS-states, a new SS-state is also formed: the supertriangle rotation makes superhexagon $2$ compatible, resulting in the appearance of the localized loop SS-state $\vec{\sss}^{C}_2$ (see Fig.~\ref{fig_defects}f). In general, in a network denoted $X$, the two SS-states $\vec{\sss}^{X}_{i}$ and $\vec{\sss}^{X}_{j}$---with nonzero stress on the bond $r$ that is removed due to a supertriangle rotation---are recombined to form a new SS-state $\vec{\sss}^{X+1}_{ij}$. This SS-state is found via the equation
	\begin{equation}\label{eq:deloc_sss}
	\vec{\sss}^{X+1}_{ij} =  \vec{\sss}^{X}_{i} - \frac{\vec{\sss}^{X}_{i}\cdot \unitvec{r}}{\vec{\sss}^{X}_{j}\cdot \unitvec{r}}\vec{\sss}^{X}_{j},
	\end{equation}
	where $\unitvec{r}$ is a bond stress vector with unity value on bond $r$, and zero value on all other network bonds.\\
	
	Finally, rotating a last supertriangle in network $C$ produces network $D$ that contains a single topological defect; the number of SS-states remains the same. The delocalized state $\vec{\sss}^{C}_{123}$, with its nonzero stress on the removed bond $r$, is no longer an SS-state; however, the loop SS-state $\vec{\sss}^{C}_2$ is retained, and a new loop SS-state $\vec{\sss}^{D}_3$ arises in the newly formed compatible superhexagon (see Fig.~\ref{fig_defects}h). Note that the SS-states of network $D$, with its single incompatible superhexagon, can be identified directly. Since this network is incompatible, it has $2H-1$ SS-states; $H$ of these are radial SS-states that are localized in all superhexagons, and $H-1$ SS-states are localized on the $H-1$ compatible superhexagons.\\
	
	In general, a complete basis of SS-space can be obtained for any $H$-superhexagon incompatible metamaterial with $H_o>1$ odd loops (see Appendix~\ref{app:delocal_sss}) by constructing the $H_o-1$ delocalized SS-states (Sec.~\ref{sec:sss}) via the steps shown in Fig.~\ref{fig_defects}e-g. Thus, an independent, yet non-unique and non-orthogonal basis of SS-space can be constructed in each of our mechanical metamaterials.\\
	
	This procedure illustrates that in all cases, whether the metamaterial contains no, one, or more local odd loops, the SS-space is spanned by a complete basis consisting of radial SS-states; loop SS-states localized in compatible superhexagons; and delocalized linear combinations of loop SS-states running between incompatible superhexagons. Such extended SS-states are reminiscent of flux lines that connect pairs of defects in artificial spin-ice models~\cite{Nascimento2012}.\\
	
	\subsection{Process I: supertriangle rotation from a compatible to an incompatible geometry}\label{subsec:response_evolution_process_I}
	
	\begin{figure*}
	\centering
	\includegraphics[]{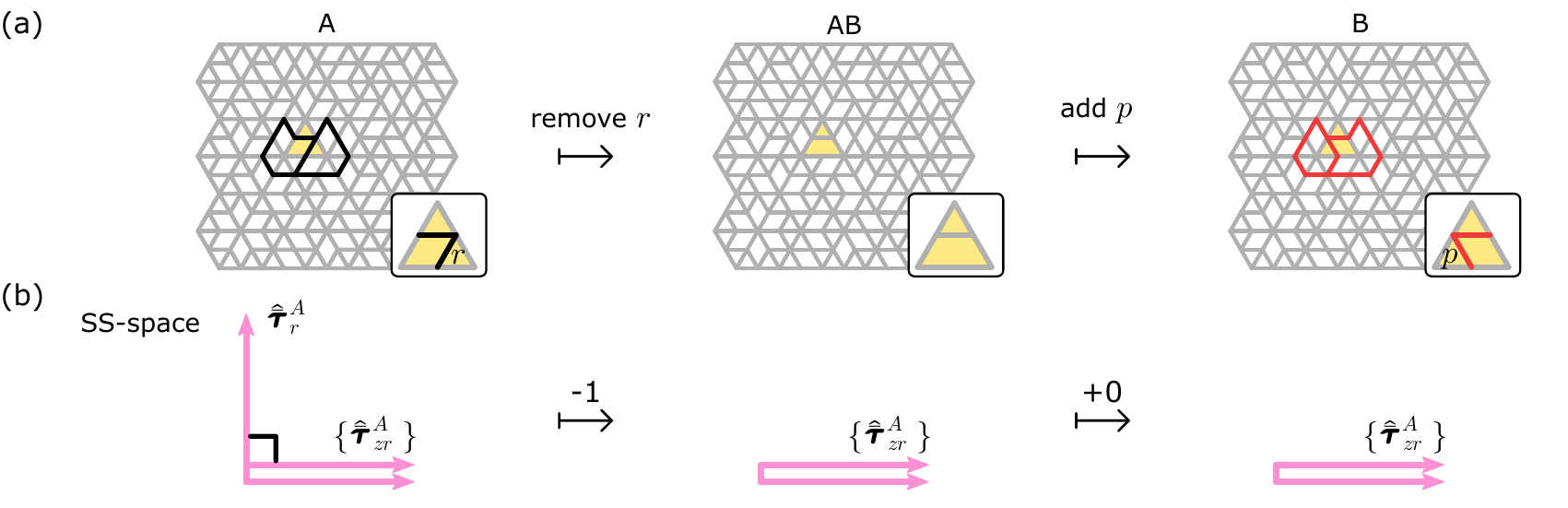}
	\caption{
		Evolution of the SS-space under a supertriangle rotation according to process I.
		(a) A compatible network $A$ is transformed to an incompatible network $B$ via an intermediate network $AB$, by first removing bond $r$ and then adding bond $p$ (insets).
		(b) For network $A$, we construct orthogonal bases for the SS-space that contain the states $\{ \unitvec{\bar{\sss}}_{zr}^A\}$ that have zero stress on bond $r$ and that remain in the SS-spaces of network $AB$ as well as $B$. The full basis of network $A$ additionally contains an SS-state $\unitvec{\bar{\sss}}_r^A$ that is removed during the architectural transformation (see text). Black square signifies orthogonality, and arrows with numbers indicate changes in the dimensions of the SS-space.
	}
	\label{fig_process_I}
	\end{figure*}	
	
	Now that we are able to construct bases of the SS-spaces of our metamaterials, we are in a position to understand how the SS-spaces change under architectural transformations, beginning with process I that converts a compatible to an incompatible metamaterial.\\
	
	\textit{(i)} We first construct a suitable orthogonal basis for the SS-space for a compatible network $A$.
	Our goal is to identify the unique SS-state in network $A$, $\vec{\bar{\sss}}_r^A$, that has a finite stress on bond $r$ and that therefore is not present in network $AB$; and to construct the set of $2H-1$ orthonormal basis vectors $\{\vec{\bar{\tau}}_{zr}^A\}$ that have zero stress on bond $r$, are perpendicular to $\vec{\bar{\sss}}_r^A$, and remain present in network $AB$. Here, the symbol $\vec{\sss}$ indicates an SS-state; the superscript $A$ indicates the network; and the subscripts $r$ or $zr$ indicate whether the vector has nonzero or zero stress on bond $r$, respectively.\\
	
	We construct $\vec{\bar{\sss}}_r^A$ and $\{\vec{\bar{\tau}}_{zr}^A\}$ as follows, as shown in Fig.~\ref{fig_sssortho}. First, as bond $r$ is shared between exactly two even local loops in $A$ (Fig.~\ref{fig_sssortho}a), there are two unique loop SS-states $\vec{\sss}_1^{A}$ and $\vec{\sss}_2^{A}$ with nonzero stress on $r$ (Fig.~\ref{fig_sssortho}b), and $2H-2$ loop SS-states $\{\vec{\sss}_i^{A}\}_{i=3}^{2H-2}$ with zero stress on $r$. We construct an additional SS-state with zero stress on $r$ by taking a linear combination of $\vec{\sss}_1^{A}$ and $\vec{\sss}_2^{A}$ (Fig.~\ref{fig_sssortho}c):
	\begin{equation}\label{eq:unique_sss_1}
	\vec{\sss}_{12}^{A} = \vec{\sss}_1^{A} - \frac{\vec{\sss}_1^{A}\cdot\unitvec{r}}{\vec{\sss}_2^{A}\cdot\unitvec{r}}~\vec{\sss}_2^{A}~,
	\end{equation}
	where $\unitvec{r}$ is the unit bond stress vector with unity value on bond $r$, and zero stress elsewhere. The SS-state $\vec{\sss}_{1}^{A}$ is, by construction, the only state in our SS-space basis $\{\vec{\sss}_1^{A},\vec{\sss}_{12}^{A},\{\vec{\sss}_i^{A}\}_{i=3}^{2H}\}$ with nonzero stress on $r$. We now perform a sequential Gram-Schmidt process ($\mathrm{GS}$) on the ordered set (left to right)  of SS-states to orthonormalize the basis:
	\begin{equation}\label{eq:gram_schmidt}
	\{\{\unitvec{\bar{\tau}}_{zr}^A\} ,\unitvec{\bar{\sss}}_{r}^A\}
	=
	\mathrm{GS}[\{ \{\vec{\sss}_i^{A}\}_{i=3}^{2H}, \vec{\sss}_{12}^{A},\vec{\sss}_1^{A} \}]~,
	\end{equation}
	where the bar and hat in $\unitvec{\bar{\sss}}$ indicate orthogonality and normality respectively. The first two SS-states of the basis are illustrated in Fig.~\ref{fig_sssortho}d. Going from network $A$ to $AB$ by removing bond $r$ removes one SS-state, which must be $\unitvec{\bar{\sss}}_r^A$ (Fig.~\ref{fig_sssortho}d), while the remaining	$\{\unitvec{\bar{\tau}}_{zr}^A\}$ span the SS-space of network $AB$. Going from network $AB$ to $B$ by
	adding bond $p$ leaves the SS-space unaffected.\\
	
	For completeness, the evolution of the complementary LB-space is presented in Appendix~\ref{app:stress_space_evolution} via a similar strategy.\\

	In summary, when a compatible metamaterial $A$ is converted to an incompatible architecture $B$ according to process I, the evolution of the SS-space is simple once an appropriate basis is constructed. The SS-spaces of architecturally related networks $A$ and $B$ are identical up to the SS-state $\unitvec{\bar{\sss}}_{r}^{A}$, present in network $A$, but not in $B$, as illustrated schematically in Fig.~\ref{fig_process_I}.\\

	\begin{figure*}
		\centering
		\includegraphics[]{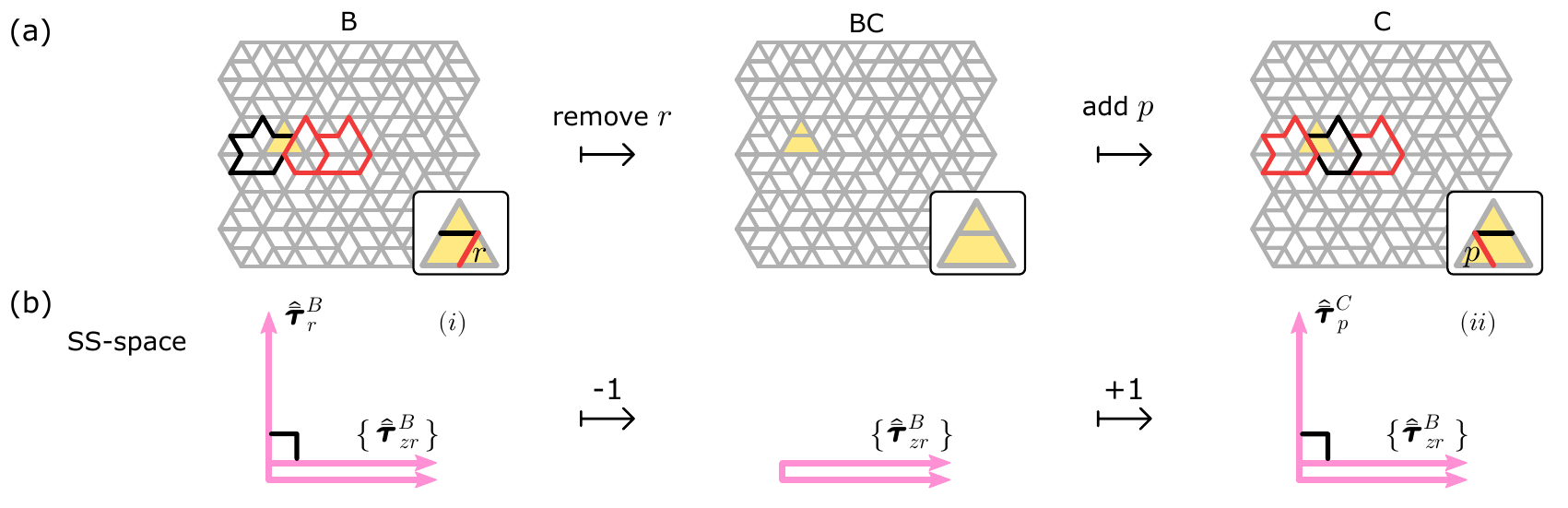}
		\caption{
			Evolution of the SS-space under a supertriangle rotation according to process II.
			(a) An incompatible network $B$ is transformed to an incompatible network $C$ via an intermediate network $BC$, by first removing bond $r$ and then adding bond $p$ (insets).
			(b) For network $B$, we construct orthogonal bases for the SS-space that contain the states $\{ \unitvec{\bar{\sss}}_{zr}^B\}$ that remain in the SS-space of network $BC$ as well as in that of $C$. The full bases of networks $B$ and $C$ additionally contain the respective SS-states $\unitvec{\bar{\sss}}_r^B$ and $\unitvec{\bar{\sss}}_r^C$ that are removed and added during the architectural transformation (see text). Black squares signify orthogonality, and arrows with numbers indicate changes in the dimensions of the SS-space.
		}
		\label{fig_process_II}
	\end{figure*}

	\subsection{Process II: supertriangle rotation from an incompatible to another incompatible geometry}\label{subsec:response_evolution_process_II}
	
	We now discuss the stress space changes of process II, converting an incompatible network $B$ to an intermediate network $BC$ and finally to a distinct incompatible network $C$, as shown in Fig.~\ref{fig_process_II}a. There are two calculations necessary to understand process II, and they are shown schematically in Fig.~\ref{fig_process_II}b. With minor modifications, these calculations follow the strategy developed for process I above. We again denote the removed and added bonds by $r$ and $p$, although we note that these refer to different bonds than in process I.\\
	
	\textit{(i)} We construct an orthogonal basis for the SS-space of the incompatible network $B$ by identifying its unique SS-state, $\unitvec{\bar{\sss}}_r^B$, that has a finite stress on bond $r$ (and is thus not present in network $BC$), and  constructing the remaining set of orthogonal basis vectors $\{\unitvec{\bar{\sss}}_{zr}^B\}$ that have zero stress on bond $r$ (and thus remain present in network $BC$). To do this, we use the same method as for process I, step \textit{(i)} above: we first construct $\tau_{12}^B$, create a basis $\{\vec{\sss}_1^{B},\vec{\sss}_{12}^{B},\{\vec{\sss}_i^{B}\}_{i=3}^{2H}\}$, and perform a sequential Gram-Schmidt process (Eq.~(\ref{eq:gram_schmidt})) to obtain the orthogonal basis $ \{ \{\unitvec{\bar{\sss}}_{zr}^B\}, \unitvec{\bar{\sss}}_r^B \}$. Going from network $B$ to $BC$ by removing bond $r$, the SS-state $\unitvec{\bar{\sss}}_r^B$ is removed from the SS-space (see Fig.~\ref{fig_process_II}b, left).\\
	
	\textit{(ii)} To go from network $BC$ to network $C$, we add bond $p$, which increases the dimension of the SS-space by one. To construct a basis for the new SS-space, we use an inverse procedure and start from network $C$, constructing a basis suitable for removing bond $p$ to obtain network $BC$. We use the same procedure as in step \textit{(i)} above, and we readily obtain a basis $\{ \{\unitvec{\bar{\sss}}_{zp}^C\}, \unitvec{\bar{\sss}}_{p}^C \}$. Noting that removing bond $p$ from network $C$ and removing bond $r$ from network $B$ produces the same network $BC$, it trivially follows that $\{\unitvec{\bar{\sss}}_{zp}^C\}$ = $\{\unitvec{\bar{\sss}}_{zr}^B\}$. Hence, the step from network $BC$ to $C$ simply adds the basis vector $\unitvec{\bar{\sss}}_{p}^C$ to the SS-space (see Fig.~\ref{fig_process_II}b, right).\\
	
	For completeness, the evolution of the complementary LB-space is presented in Appendix~\ref{app:stress_space_evolution} following a similar set of calculations.
	
	Together, steps \textit{(i)} and \textit{(ii)} describe the evolution of the SS-space for process II, converting an incompatible network $B$ to a second, distinct incompatible network $C$. The SS-spaces of architecturally related networks $B$ and $C$ are identical up to the SS-state $\unitvec{\bar{\sss}}_{r}^{B}$, present in network $B$, but not in $C$; and the SS-state $\unitvec{\bar{\sss}}_{p}^{C}$, present in network $C$, but not in $B$.
	
	\subsection{Mechanical interpretation and consequences}\label{subsec:response_evolution_interpretation}
	
	\begin{figure*}
		\centering
		\includegraphics[]{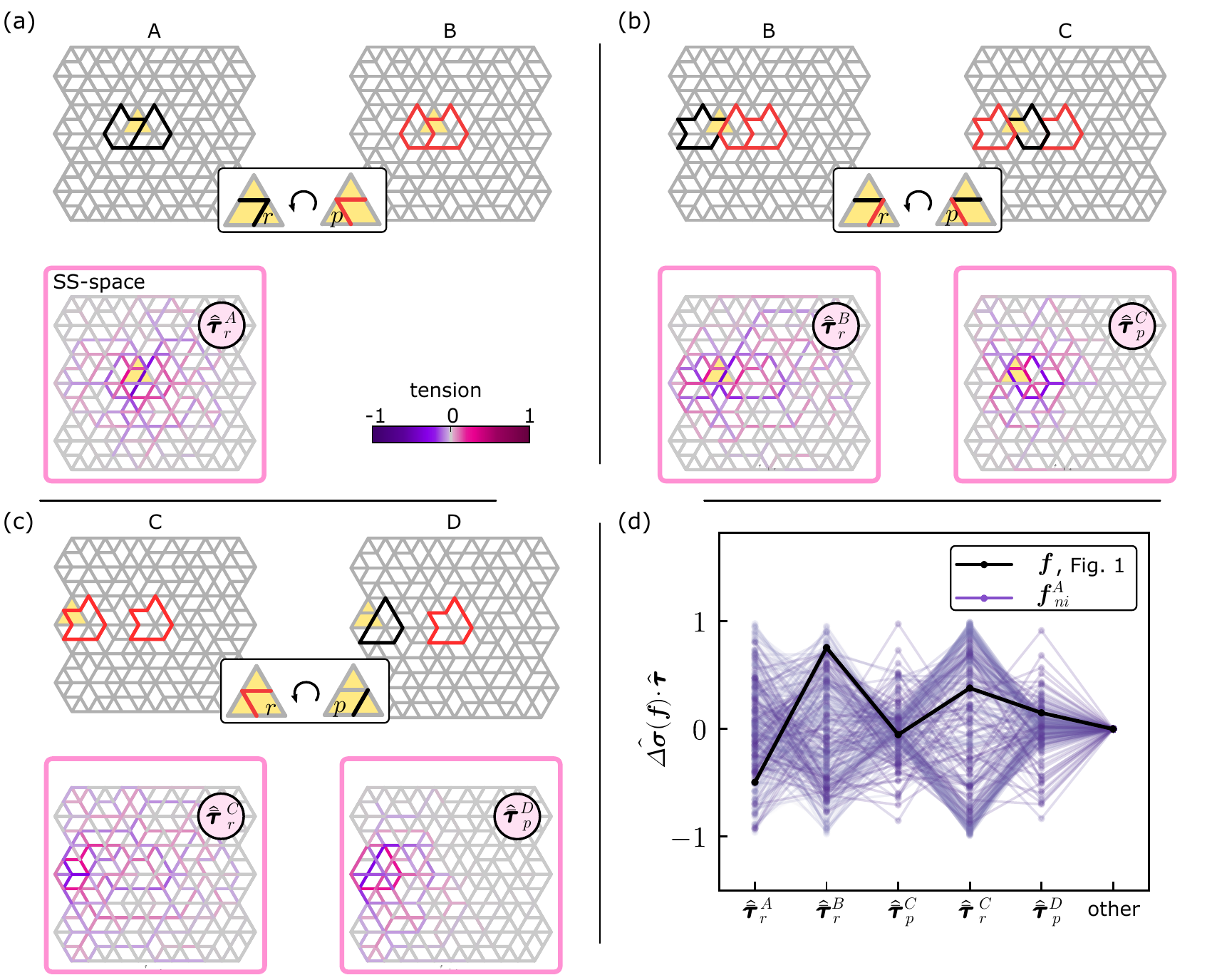}
		\caption{
			Examples of the reconfiguration of a metamaterial's SS-states under a sequence of supertriangle rotations, shown for network pairs $A$--$B$, $B$--$C$, and $C$--$D$.
			(a) A compatible network $A$ (left) is transformed to exhibit a structural defect in network $B$ (right) by rotating a supertriangle, effectively removing bond $r$ and adding bond $p$ (inset). Local loops whose parity is modified are indicated cf. Fig.~\ref{fig_localsss}. The unique SS-state $\unitvec{\bar{\sss}}_1^A$ with nonzero stress on bond $r$ that is not an SS-state of network $B$ is shown.
			(b) Network $B$ is transformed into network $C$, which contains two topological defects. The evolution of the SS-space is set by the two mutually exclusive SS-states $\unitvec{\bar{\sss}}_r^B$ and $\unitvec{\bar{\sss}}_p^C$.
			(c) Network $C$ is converted to network $D$ containing a single topological defect. The SS-space is modified such that only the two SS-states $\unitvec{\bar{\sss}}_r^C$ and $\unitvec{\bar{\sss}}_p^D$ are not shared by the two networks.
			(d) The stress response difference $\Delta\vec{\sigma}$ between networks $A$ and $D$ under identical loading is a linear combination of the five mutually exclusive SS-states. $\Delta\vec{\sigma}$ is calculated for all independent normal mode loads $\vec{f}^A_{ni}$ of network $A$ (see text), as well as the load illustrated in Fig.~\ref{fig_intro}b. The overlap of the normalized stress difference with the five SS-states is shown; it has no component outside of their span.
		}
		\label{fig_spacechanges}
	\end{figure*}
	
	The above results show how the SS-space changes under a supertriangle rotation. Specifically, we constructed the mutually exclusive SS-states of two architecturally related networks. There is one such SS-state for a network pair where the dimension of the LB-space changes (process I), two such SS-states for  networks where the dimension of the LB-space does not change (process II), and no such SS-states for process III.\\
	
	Due to the linear-algebraic structure of our model, we have argued that the SS-space evolution between two architecturally related metamaterials governs their difference in stress response. After all, the stress response of both metamaterials must be perpendicular to their respective SS-spaces. This enables us to answer the following question: \textit{when two metamaterials with distinct architectures are subjected to the same external nodal load $\vec{f}$, what is the difference $\Delta\vec{\sigma}$ in their stress response?}\\
	
	We show an explicit example for the three network pairs $A-B$ and $B-C$ in Fig.~\ref{fig_spacechanges}a--b, corresponding to processes I and II respectively. The figure illustrates the SS-states that mutate under architectural transformations. When network $A$ is transformed into network $B$, the only difference between the two respective SS-spaces is the SS-state $\unitvec{\bar{\sss}}_r^A$ (Fig.~\ref{fig_spacechanges}a, bottom). Thus, the stress difference between networks $A$ and $B$ under identical supported loading is parallel to $\unitvec{\bar{\sss}}_r^A$. To show this precisely, some linear algebra is necessary; details are shown in Appendix~\ref{app:interpretation}. With this result, we can understand the localization of the stress response difference between networks $A$ and $B$, introduced in Fig.~\ref{fig_intro}a: the localization of the stress response difference is due to the localization of the SS-state $\unitvec{\bar{\sss}}_r^A$ around the removed bond $r$.\\
		
	Similarly, the stress response difference between the networks $B$ and $C$, related via process II, is spanned by the changed SS-states $\unitvec{\bar{\sss}}_r^B$ and $\unitvec{\bar{\sss}}_p^C$ (Fig.~\ref{fig_spacechanges}b, bottom; see Appendix~\ref{app:interpretation} for details).\\
	
	As a consequence, we can make an inductive statement about the stress response difference between a pair of networks related by multiple, consecutive block rotations, such as the network pair $A-D$ shown in Fig.~\ref{fig_intro}b. The stress response difference between the two networks must be limited to the span of SS-states that have changed during the sequential transformations. The network with a topological defect ($D$) is related to the compatible network ($A$) by a minimal number of three architectural transformations, shown in Fig.~\ref{fig_spacechanges}a--c, that correspond to processes I, II, and II respectively. As a consequence, the stress response difference between networks $A$ and $D$ should be contained in a five-dimensional stress subspace of changed SS-states (Fig.~\ref{fig_spacechanges}a--c, bottom). To confirm this, we calculate the stress response difference between networks $A$ and $D$ under all $N_b - N_{SS}$ independent supported loads of network $A$. We choose the independent supported loads to be the supported normal loads $\vec{f}^A_{ni}$ (i.e. left singular vectors with nonzero singular values of the kinematic matrix of network $A$). The overlap of the resulting normalized stress response differences $\hat{\Delta\vec{\lbs}}$ with the five normalized SS-states is shown in Fig.~\ref{fig_spacechanges}d. The data demonstrate that the stress response difference is a linear combination of only the five mutually exclusive SS-states for any applied load, with zero projection on any other stress states. Results are also shown for the particular stress response difference under the loading illustrated in Fig.~\ref{fig_intro}b (right). Thus, the stress response difference shown in Fig.~\ref{fig_spacechanges}d is confirmed to be a linear combination of the five SS-states, each of which is concentrated in a different part of the network. Since the stress response difference is a linear combination of mutated SS-states with different localizations, the total stress response difference is diffuse.\\

\section{Re-steering a stress response with architectural transformations}\label{sec:design}

\begin{figure*}
	\centering
	\includegraphics[]{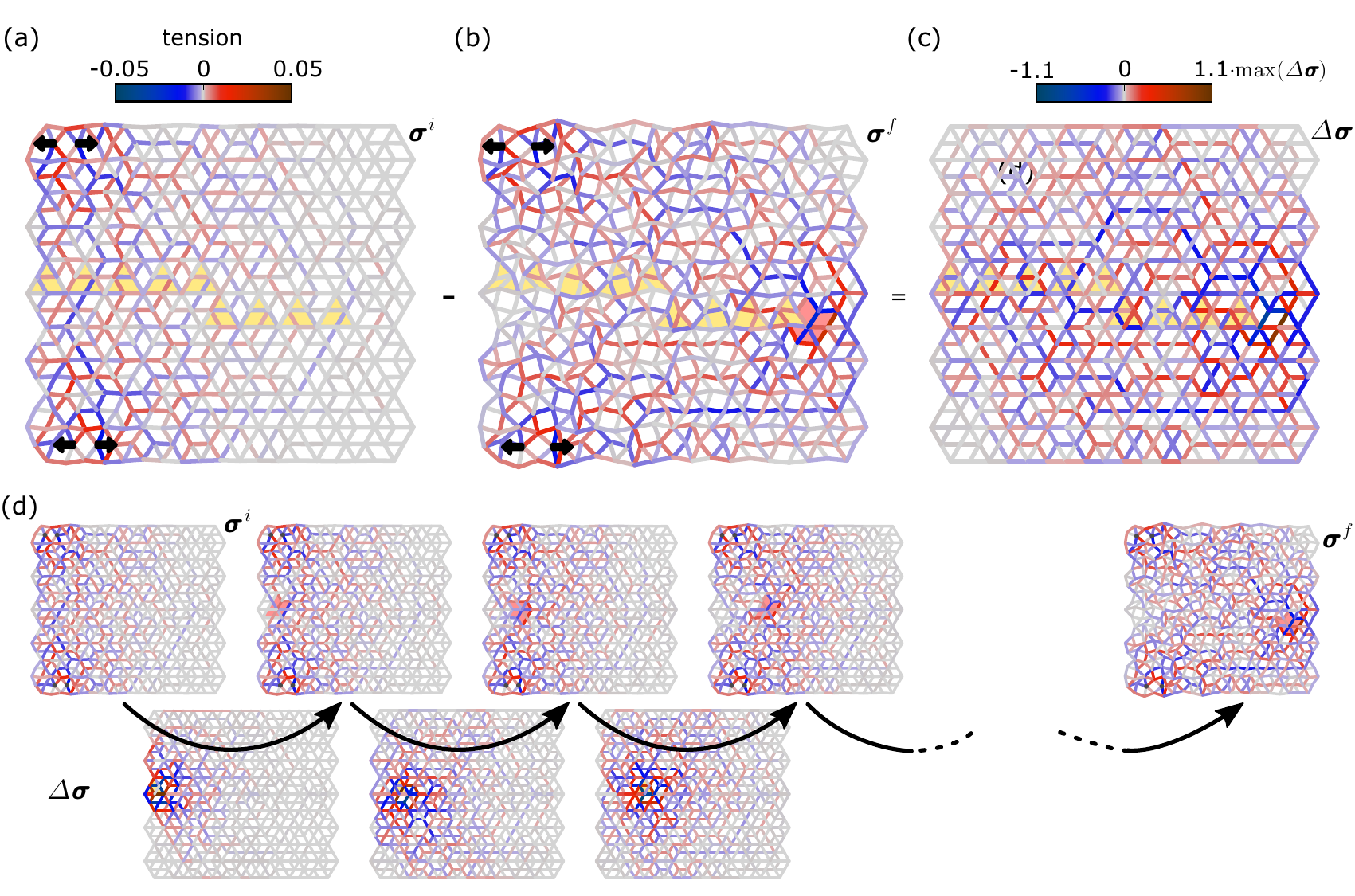}
	\caption{(a) An initially compatible metamaterial under loading at the network's leftmost corners (arrows, length multiplied by a factor 100 for clarity) concentrates stresses $\vec{\lbs}^i$ (colours) along the shortest path between the two probing points. Nine sequential supertriangle rotations (yellow triangles) introduce a topological defect from the left boundary and guide it to the right.
		(b) Once the topological defect has been moved to the right boundary, the stress field $\vec{\lbs}^f$ is diverted to run between the two probing points and along the right side of the topological defect (odd local loop highlighted with red infill).
		(c) The differential stress response  $\Delta\vec{\lbs}$ of the two networks is such that stresses on the left of the system are decreased, while stresses on the right increase. $\Delta\vec{\lbs}$ is a linear combination of the 17 SS-states that have changed during the nine sequential architectural transformations. 
		(d) The stress response and stepwise stress response difference for the first three intermediate steps is shown. Intermediate stress response difference are linear combinations of SS-states that are quasilocalized near the rotated supertriangles. The SS-states produce a typical stress re-steering that affects stress magnitudes near the moving topological defect: stresses to the left are decreased, while stresses on the right increase.
	}
	\label{fig_conclusions}
\end{figure*}

In this section, we show that our understanding of SS-space modifications during architectural transformations allows us to explain how the inclusion of a topological defects affects the stress response field of a metamaterial.\\

In previous work, we have shown that metamaterials containing a single topological defect show unusual stress-localizing behaviour when compared to a compatible metamaterial~\cite{Hecke2019}. Specifically, consider a compatible network; an example of a large compatible network containing 95 superhexagons is shown in Fig.~\ref{fig_conclusions}. We pick two supertriangles at the left top and bottom corners for actuation. To make sure that we have a supported load, and for simplicity, we force both supertriangles with load dipoles that actuate their local FM, but that is not compatible with the network's global FM and is therefore a supported load. Under this driving, stresses are concentrated along the leftmost sample edge, running along the shortest path between the two actuation points (Fig.~\ref{fig_conclusions}a). When the metamaterial undergoes a particular sequence of supertriangle rotations to generate a topological defect that progressively moves from left to right through the system, the same loading conditions produce a stress field that runs along the rightmost edge of the network instead (Fig.~\ref{fig_conclusions}b). The differential stress response is concentrated on the right side of the system (Fig.~\ref{fig_conclusions}c). Based on the evolution of SS-space during each supertriangle rotation, we can understand why this unusual stress-localizing behaviour takes place.\\

Starting from the compatible structure, we rotate a supertriangle at the leftmost edge to locally create a topological defect. This removes a SS-state at the leftmost edge of the system (Fig.~\ref{fig_conclusions}d, left). The particular removed SS-state is structured so that the stress response of the new network is reduced at the left and increased to the right of the newly created topological defect. In the next transformation step, we shift the topological defect to the right by rotating a supertriangle on the right side of the topological defect. This transformation locally modifies the SS-states, which are again configured such that the stress response is decreased to the left and increased to the right, so that stresses are steered along the right edge of the topological defect. Repeating this process leads to the path of highest stress concentration to be pushed farther and farther towards the right side of the system, ahead of the direction of `motion' of the topological defect (Fig.~\ref{fig_conclusions}d, middle). Finally, after the transformation sequence is complete, the topological defect is located at the rightmost side of the network; the stress field runs between the two actuation points around the defect along the right edge, leaving the left edge with a lowered stress response (Fig.~\ref{fig_conclusions}d, right). SS-states that are modified during such transformations fully determine the difference in stress response under an equal applied load.\\

\section{Conclusions and outlook}\label{sec:outlook}
In previous work, SS-states have been used  to design localized mechanical responses in materials with a topologically nontrivial band structure~\cite{Kane2013,Paulose2015a,Chen2016,Liu2018,Rocklin2016}, or to investigate the structure and mechanical response of mechanical networks~\cite{Paulose2015,Lerner2018,Guest2005} and jammed particle packings~\cite{Sussman2016,Hexner2018a,Hexner2018,Ji2019,Wijtmans2017,Bassett2015,Snoeijer2004a,Ramola2017,Lois2009,Lubensky2015}. In contrast, here we have worked out in detail
how architectural {\em transformations} govern the {\em evolution} of the SS-states, LB-states, and mechanical response of a complex mechanical metamaterial \cite{Hecke2019}.\\

In particular, we started from a linear-algebraic description of network mechanics, which dictates that the stress difference of architecturally related networks under identical loading is governed by the networks' differing SS-spaces. It should be noted here that this result holds not only for the metamaterial architectures presented in this work, but for \textit{any} network material whose architecture is transformed by removing a bond, and then adding a bond at another position: under identical supported loads, the response difference between the two architecturally related networks is governed by their mutually exclusive SS-states.\\

For the specific family of metamaterials considered here, closed-form SS-states spanning the full SS-space were constructed straightforwardly, due to the regular geometry of the metamaterial building block. We then considered rotations of a single triangular building block as the fundamental architectural transformations that can introduce (topological) defects into formerly compatible designs~\cite{Hecke2019}. These rotations were shown to lead to distortions of the SS-space that we calculated explicitly. In turn, since changes in the SS-space govern the evolution of the metamaterial's stress response under externally applied loads, we were able to explicitly calculate how the response of a metamaterial evolves under architectural transformations. Finally, we demonstrated how these insights clarify how topological defects steer stress fields.\\

While our approach helps understand the steering of stresses in the particular case of a moving topological defect, designing a target stress response with an inverse procedure is more complex. Suppose, for example, that we aim to construct a sequence of architectural transformations to generate a given target stress response, starting from a particular metamaterial design and loading conditions. In general, this requires an in-depth analysis of the evolution of the SS-states to ensure their cumulative contribution leads to the desired stress response. Nevertheless, our approach suggests a systematic pathway to do so. Moreover, metamaterial designs may be constructed where the SS-states are a priori known or more easy to construct, simplifying the practical implementation of our approach to design the (differential) stress response of complex metamaterials.\\

\section{Acknowledgements}
We thank Aparna Baskaran, Roni Ilan, Edan Lerner, Jayson Paulose, Ben Pisanty, and Eial Teomy for fruitful discussions. This research was supported in part by the Israel Science Foundation Grant No. 968/16, and by the Israeli Ministry of Science and Technology.

\clearpage

\setcounter{figure}{0}

\appendix
\renewcommand{\thefigure}{S\arabic{figure}}

\maketitle

\section{Constructing delocalized SS-states}\label{app:delocal_sss}

\begin{figure*}
	\centering
	\includegraphics[]{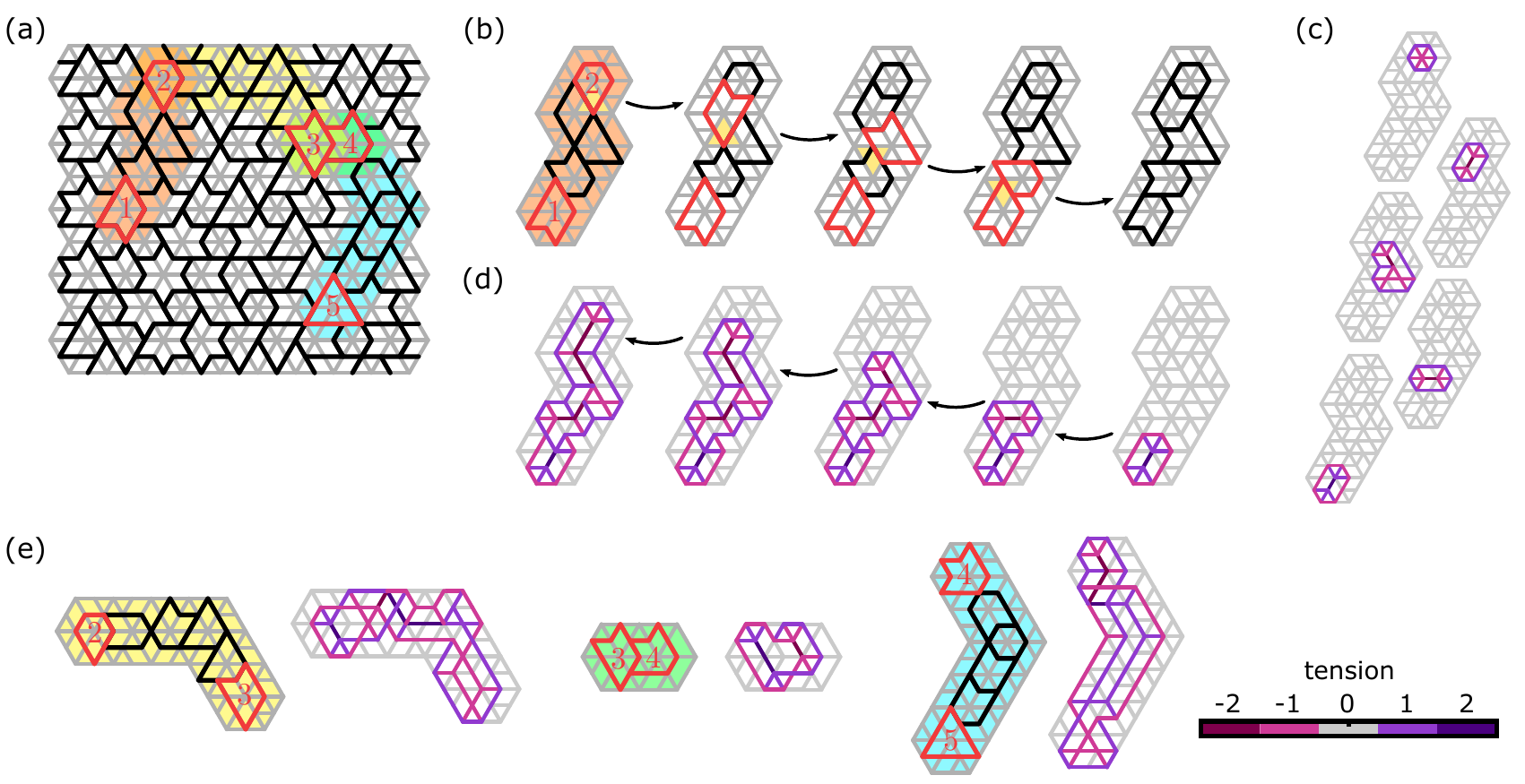}
	\caption{(a) An incompatible metamaterial architecture containing $H_o=5$ odd local loops (numbers $1$-$5$, red bold lines) and $H-H_0=72$ even ones (black bold lines). The metamaterial contains $H_o-1=4$ delocalized SS-states, which are constructed on metamaterial paths connecting four independent pairs of incompatible hexagons (infills in orange, 1-2; yellow, 2-3; green, 3-4; and blue, 4-5).
		(b) The incompatible metamaterial strip between superhexagons 1,2 is made compatible by sequential supertriangle rotations (yellow triangles, arrows) that change the parity of local loops.
		(c) The compatible metamaterial's SS-states are spanned by radial (not shown) and loop SS-states (colour bar).
		(d) The loop SS-states are recombined into a delocalized SS-state of the incompatible metamaterial strip using Eq.~(\ref{eq:deloc_sss}) (arrows), yielding a delocalized SS-state between the incompatible superhexagons 1 and 2.
		(e) With the procedure demonstrated in b---d, the other three delocalized SS-states are constructed between the remaining pairs of odd local loops.
	}
	\label{fig_SI_delocsss}
\end{figure*}

We show how to construct the $H_o-1$ delocalized SS-states for any $H$-superhexagon metamaterial with $H_o>1$ odd local loops. We consider the schematic shown in Fig.~\ref{fig_SI_delocsss}, which illustrates how delocalized SS-states can be constructed iteratively. The network shown contains $H_o=5$ odd local loops (numbered $1$-$5$) that contains $H_o-1=4$ delocalized SS-states (Fig.~\ref{fig_SI_delocsss}a).\\

We first show how to create a delocalized SS-state running between a pair of two odd local loops (numbered $1$, $2$). We start by identifying a small subsection of the network to construct the SS-state in, consisting of the two incompatible superhexagons containing the odd local loops, and an arbitrary string of compatible superhexagons that connects the pair (Fig.~\ref{fig_SI_delocsss}a, orange infill). We then transform this metamaterial strip into a compatible structure---in which all SS-states are known exactly---via a series of supertriangle rotations (Fig.~\ref{fig_SI_delocsss}b, yellow triangles, arrows) that sequentially flip the parity of the local loops. We are left with a compatible structure in which all loop and radial SS-states are found by inspection (Fig.~\ref{fig_SI_delocsss}c, radial SS-states not shown for clarity). As explained in Fig.~\ref{fig_defects}e-g, these loop SS-states may then be recombined via sequential application of Eq.~(\ref{eq:deloc_sss}) under inversion of the applied supertriangle rotations, analogous to the construction discussed in Sec.~\ref{subsec:response_evolution_sss}. The linear combination of loop SS-states thus produces a delocalized SS-state of the metamaterial strip with the two odd local loops $1$ and $2$ (Fig.~\ref{fig_SI_delocsss}d, arrows).\\

In a metamaterial with $H_o$ odd loops, we can find $H_o-1$ independent delocalized states using the above procedure. Independence is ensured by selecting $H_o-1$ independent pairs of incompatible superhexagons (such that each is selected at least once), with strings of compatible superhexagons running between them. Figure~\ref{fig_SI_delocsss}e demonstrates the three remaining delocalized SS-states found between defect pairs $(2,3)$, $(3,4)$, and $(4,5)$ in our example.\\

It should be noted that the delocalized states are not unique: their shape depends on the path between each defect pair, and the choice of supertriangle rotations. However, the space spanned by the resulting basis of SS-states does not depend on the path choice. In particular, this procedure renders an independent, non-orthogonal set of $H_o-1$ delocalized SS-states. Together with the known radial and loop SS-states, which are identified by inspection, a complete and independent basis of SS-space can be found for our metamaterials with any defect configuration.

\section{Evolution of LB-spaces under architectural transformations}\label{app:stress_space_evolution}

\begin{figure*}
	\centering
	\includegraphics[]{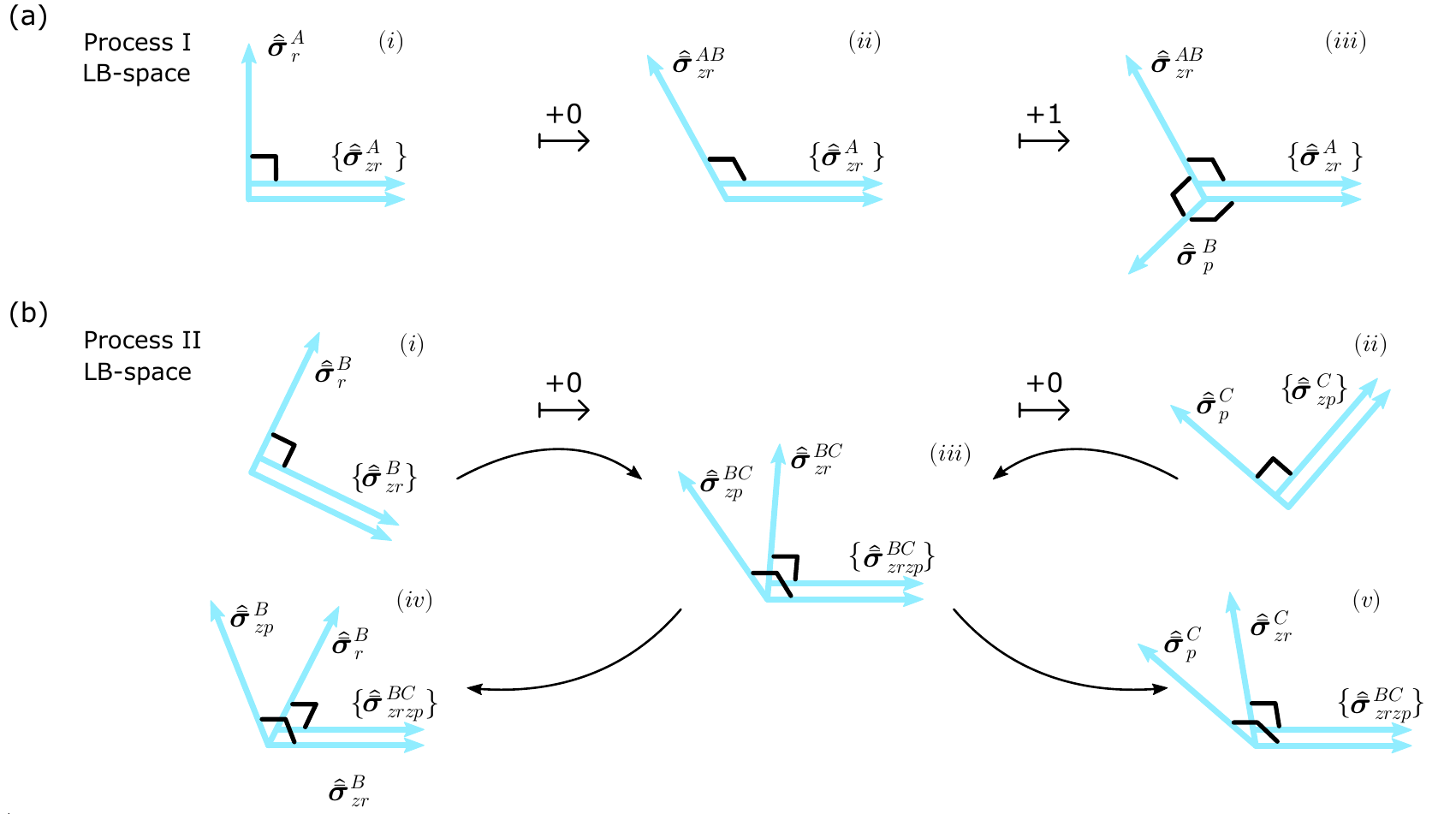}
	\caption{
		Evolution of the LB-space under a supertriangle rotation according to processes I and II.
		(a) Process I: a compatible network $A$ is transformed to an incompatible network $B$ via an intermediate network $AB$, by first removing bond $r$ and then adding bond $p$ (see Fig.~\ref{fig_process_I}). For network $A$, we construct an orthogonal basis for the LB-space that contains those states $\{ \unitvec{\bar{\lbs}}_{zr}^A\}$ that remain in network $AB$ as well as $B$. The full bases of networks $A$ and $B$ additionally contain a state that is added ($\unitvec{\bar{\lbs}}_p^B$) and a state that is modified ($\unitvec{\bar{\lbs}}_r^A$ to $\unitvec{\bar{\lbs}}_{zr}^{AB}$) during the architectural transformation; for details on the execution of steps \textit{(i)}-\textit{(iii)}, see text.
		(b)Process II: an incompatible network $B$ is transformed to an incompatible network $C$ via an intermediate network $BC$, by first removing bond $r$ and then adding bond $p$ (see Fig.~\ref{fig_process_II}). For network $B$ ($P$), we construct an orthogonal basis for the LB-space that contains states $\{ \unitvec{\bar{\lbs}}_{zr}^B\}$ ($\{ \unitvec{\bar{\lbs}}_{zp}^C\}$) without stress on $r$ ($p$), and $\unitvec{\bar{\lbs}}_{r}^B$ ($\unitvec{\bar{\lbs}}_{p}^C$) with finite stress there. We construct a suitable basis of LB-space for the intermediate network $BC$ (with no stress on $p$ or $r$) via an orthogonalization procedure, which produces LB-states $\{  \unitvec{\bar{\lbs}}_{zrzp}^{BC} \}$ that are shared with networks $B$ and $C$, and states $\unitvec{\bar{\lbs}}_{zp}^{BC}, \unitvec{\bar{\lbs}}_{zr}^{BC}  $ that are modified to states $\unitvec{\bar{\lbs}}_{zp}^{B}$ and $\unitvec{\bar{\lbs}}_{zr}^{C} $ in networks $B$ and $C$ respectively. For details on the execution of steps \textit{(i)}-\textit{(v)}, see text. Black squares signify orthogonality, and arrows with numbers indicate changes in the dimensions of the LB-space.
	}
	\label{fig_SI_processI_II}
\end{figure*}

In Sec.~\ref{sec:response_evolution}, we discussed the evolution of a metamaterial's stress space (consisting of the SS- and complementary LB-space) under architectural transformations. We demonstrated that the evolution of the SS-space is limited to one, two, or no changing SS-states for distinct types of supertriangle rotations, denoted process I, process II, and process III, respectively. Here, we derive the concomitant evolution of the metamaterial's LB-space for all three processes.

\subsection{Process I: compatible to incompatible metamaterial}
We now describe the evolution of the LB-space when a compatible network $A$ is transformed into an incompatible network $B$; this evolution is shown schematically in Fig.~\ref{fig_SI_processI_II}a. The architectural transformation occurs via a supertriangle rotation that removes a bond $r$ and adds a bond $p$ (see Fig.~\ref{fig_process_I}a). The LB-space evolution is closely related to the evolution of the SS-space discussed in Sec.~\ref{subsec:response_evolution_process_I} (see Fig.~\ref{fig_process_I}b), and involves three separate calculations \textit{(i)}-\textit{(iii)} below.

\textit{(i)}
We aim to construct a basis for the LB-space of network $A$ that consists of one LB-state, $\unitvec{\bar{\lbs}}_r^A$, that has a finite stress on bond $r$, and a remaining set of orthogonal vectors $\{\unitvec{\bar{\lbs}}_{zr}^A\}$ that have zero stress om bond $r$ (Fig.~\ref{fig_process_I}c, left). Under removal of bond $r$, only the LB-state $\unitvec{\bar{\lbs}}_r^A$ will be modified. Since the set $\{\unitvec{\bar{\lbs}}_{zr}^A\}$ is unaffected by removing $r$ and adding $p$, we do not need to construct it explicitly, and focus on identifying $\unitvec{\bar{\lbs}}_r^A$ instead. To construct this unique LB-state with nonzero stress on bond $r$, note that the stress state $\unitvec{r}$ must be a linear combination of the SS-state $\unitvec{\bar{\sss}}_r^A$ (see Sec.~\ref{subsec:response_evolution_process_I}) and $\unitvec{\bar{\lbs}}_r^A$---the only two stress states with nonzero stress on $r$---and since $\unitvec{\bar{\sss}}_r^A$ and $\unitvec{\bar{\lbs}}_r^A$ are perpendicular, we find
\begin{equation}\label{eq:make_lbs}
	\unitvec{\bar{\lbs}}_{r}^{A} \propto \mathrm{Rej}(\unitvec{r},\unitvec{\bar{\sss}}_{r}^{A})~,
\end{equation}
as shown in Fig.~\ref{fig_SI_processI_II}a, left. Here, we define the vector rejection $\mathrm{Rej}(~)$ to be the complement of vector projection: $\mathrm{Proj}(\vec{u}, \vec{v}) = \frac{\vec{u}\cdot\vec{v}}{\vec{v}\cdot\vec{v}}\vec{v}$ and $\vec{u} = \mathrm{Proj}(\vec{u}, \vec{v}) + \mathrm{Rej}(\vec{u}, \vec{v})$, so that $\mathrm{Rej}(\vec{u}, \vec{v}) := \vec{u} - \frac{\vec{u}\cdot\vec{v}}{\vec{v}\cdot\vec{v}}\vec{v}$.

\textit{(ii)}
When bond $r$ is removed from network $A$, the LB-state $\unitvec{\bar{\lbs}}_r^A$ must disappear; the LB-states $\{ \unitvec{\bar{\lbs}}_{zr}^A \}$ remain. However, as the number of LB-states in $AB$ is the same as in network $A$ (see above), the intermediate network $AB$ must contain a new LB-state, $\unitvec{\bar{\lbs}}_{zr}^{AB}$, with zero stress on bond $r$. This state must be perpendicular to the SS-space spanned by $\{\unitvec{\bar{\sss}}_{zr}^A\}$, and to the LB-states $\{\unitvec{\bar{\lbs}}_{zr}^A\}$. However, $\unitvec{\bar{\lbs}}_{zr}^{AB}$ does not need to be perpendicular to the state $\unitvec{\bar{\sss}}_r^A$, so that we can construct $\unitvec{\bar{\lbs}}_{zr}^{AB}$ from the states $\unitvec{\bar{\sss}}_r^A$ and $\unitvec{r}$:
\begin{equation}\label{eq:make_lbs_2}
	\unitvec{\bar{\lbs}}_{zr}^{AB} \propto Rej(\unitvec{\bar{\sss}}_r^A, \unitvec{r})~,
\end{equation}
as shown in Fig.~\ref{fig_SI_processI_II}a, middle.

\textit{(iii)}
Finally, when network $AB$ evolves to network $B$ by adding bond $p$, a new LB-state $\unitvec{\bar{\lbs}}_p^B$ must appear. The new LB-state is perpendicular to both the SS-space spanned by $\{\unitvec{\bar{\sss}}_{zr}^A\}$ as well as the LB-space spanned by $\{ \{ \unitvec{\bar{\lbs}}_{zr}^A \}, \unitvec{\bar{\lbs}}_{zr}^{AB} \}$, and has a finite stress on bond $p$. It is easy to check that the stress state $\unitvec{p}$ uniquely satisfies these criteria: $\unitvec{\bar{\lbs}}_p^B = \unitvec{p}$ (Fig.~\ref{fig_SI_processI_II}a, right).\\

In summary, as we illustrate in Fig.~\ref{fig_SI_processI_II}a and Fig.~\ref{fig_process_I}, the stress spaces of architecturally related networks $A$ and $B$ are identical up to the following four independent vectors: the SS-state $\unitvec{\bar{\sss}}_{r}^{A}$, present in network $A$, but not in $B$; the LB-state $\unitvec{p}$, present in $B$ but not in $A$; and the LB-state $\unitvec{\bar{\lbs}}_{r}^{A}$ in network $A$ that changes to the LB-state $\unitvec{\bar{\lbs}}_{zr}^{AB}$ in network $B$. These four	independent vectors are spanned by the set $\{\unitvec{\bar{\sss}}_{r}^{A}, \unitvec{r}, \unitvec{p}\}$ consisting of the mutated SS-state and the pure stress vectors on bonds $p$ and $r$.

\subsection{Process II: incompatible to incompatible metamaterial}

We now describe the evolution of the LB-space when an incompatible network $B$ is transformed into a distinct incompatible network $C$ as shown in Fig.~\ref{fig_SI_processI_II}b, via a supertriangle rotation that removes a bond $r$ and adds a bond $p$ (see Fig.~\ref{fig_process_II}a). This evolution is closely related to the evolution of the SS-space discussed in Sec.~\ref{subsec:response_evolution_process_II} (see Fig.~\ref{fig_process_II}b), and involves five separate calculations \textit{(i)}-\textit{(v)} below.\\

We can construct the LB-spaces of networks $B$ and $C$, analogous to step \textit{(ii)} in process I. This readily yields bases \textit{(i)} $\{ \{\unitvec{\bar{\lbs}}_{zr}^B \}, \unitvec{\bar{\lbs}}_{r}^{B}\}$
and \textit{(ii)} $\{ \{\unitvec{\bar{\lbs}}_{zp}^C \}, \unitvec{\bar{\lbs}}_{p}^{C}\}$ (Fig.~\ref{fig_SI_processI_II}b, left and right). However, as the sets $\{\unitvec{\bar{\lbs}}_{zr}^B \}$ and $ \{\unitvec{\bar{\lbs}}_{zp}^C \}$ are not the same,	the bases are not suitable to compare the LB-spaces.\\

\textit{(iii)} We now construct an appropriate basis for the LB-space of network $BC$, which contains a set $\{\unitvec{\bar{\lbs}}_{zrzp}^{BC} \}$ that is shared with the LB-spaces of network $B$ and $C$ (Fig.~\ref{fig_SI_processI_II}b, middle). First, we can start from the LB-basis \textit{(i)}, remove bond $r$, and
analogous to step \textit{(ii)} of process I, obtain a basis
$\{ \{\unitvec{\bar{\lbs}}_{zr}^B \}, \unitvec{\bar{\lbs}}_{zr}^{BC}\}$. Second, starting from the LB-basis {\em(ii)} and removing bond $p$ we obtain a basis
$\{ \{\unitvec{\bar{\lbs}}_{zp}^C \}, \unitvec{\bar{\lbs}}_{zp}^{BC}\}$.
These two bases both span the LB-space of network $BC$.
We now use this to construct the appropriate
basis of the LB-space, $\{\unitvec{\bar{\lbs}}_{zp}^{BC}, \unitvec{\bar{\lbs}}_{zr}^{BC}, \{\unitvec{\bar{\lbs}}_{zpzr}^{BC}\}  \}$, so that the set $\{\unitvec{\bar{\lbs}}_{zpzr}^{BC}\}$ is shared with the LB-spaces of network $B$ and $C$.
We first perform a Gram-Schmidt process on the ordered set $\{\unitvec{\bar{\lbs}}_{zp}^{BC}, \unitvec{\bar{\lbs}}_{zr}^{BC}, \{\unitvec{\bar{\lbs}}_{zr}^{B}\}  \}$, and then
define $\{\unitvec{\bar{\lbs}}_{zpzr}^{BC}\}$ as
the last $N_b-2H-1$ vectors of the resulting orthonormal basis. To facilitate comparison with networks $B$ and $C$, we
obtain a full LB-space basis of
network $BC$ by adding the vectors
$\unitvec{\bar{\lbs}}_{zp}^{BC}$ and $\unitvec{\bar{\lbs}}_{zr}^{BC}$, so that all but the first two basis vectors are orthogonal.\\

We now obtain appropriate bases for the LB-spaces of networks $B$ and $C$ as follows (see Fig.~\ref{fig_SI_processI_II}b, left and right).\\

\textit{(iv)} We construct a basis for the LB-space of network $B$ by ensuring the orthogonality of the LB-space basis of network $BC$, $\{\unitvec{\bar{\lbs}}_{zp}^{BC}, \unitvec{\bar{\lbs}}_{zr}^{BC}, \{\unitvec{\bar{\lbs}}_{zpzr}^{BC}\}  \}$, with the SS-space of network $B$. We do this by rejecting each vector on the SS-state $\unitvec{\bar{\sss}}_{r}^B$, that is present in network $B$ but not in $BC$.
This rejection procedure results in an LB-space basis of network $B$: $\{{\unitvec{\bar{\lbs}}}_{zp}^{B}, \unitvec{\bar{\lbs}}_{r}^{B}, \{\unitvec{\bar{\lbs}}_{zpzr}^{BC}\}  \}$.\\

\textit{(v)} A similar procedure results in an analogous LB-space basis for network $C$:  $\{{\unitvec{\bar{\lbs}}}_{zr}^{C}, \unitvec{\bar{\lbs}}_{p}^{C}, \{\unitvec{\bar{\lbs}}_{zpzr}^{BC}\}  \}$.\\

In summary, as shown in Fig.~\ref{fig_SI_processI_II}b and Fig.~\ref{fig_process_II}, the stress spaces of architecturally related networks $B$ and $C$ are identical up to the following vectors: the SS-state $\unitvec{\bar{\sss}}_{r}^{B}$, present in network $B$, but not in $C$; the SS-state $\unitvec{\bar{\sss}}_{p}^{C}$, present in network $C$, but not in $B$ (see Sec.~\ref{subsec:response_evolution_process_II}); the LB-state $\unitvec{\bar{\lbs}}_{r}^{B}$ in network $B$ that changes to the LB-state $\unitvec{\bar{\lbs}}_{zr}^{C}$ in network $C$; and the  LB-state $\unitvec{\bar{\lbs}}_{p}^{C}$ in network $C$ that changes to the LB-state $\unitvec{\bar{\lbs}}_{zp}^{B}$ in network $B$. These four independent vectors are spanned by the set $\{\unitvec{\bar{\sss}}_{r}^{B}, \unitvec{\bar{\sss}}_{p}^{C}, \unitvec{r}, \unitvec{p}\}$ consisting of the mutated SS-states and the pure stress vectors on bonds $p$ and $r$.

\subsection{Process III: compatible to compatible metamaterial}

A compatible network $A$ may be transformed to a distinct compatible network $A'$ by some supertriangle rotations that remove a bond $r$ and add a bond $p$. Only supertriangle rotations at the system's edge that do not change the parity of any local loops (see Sec.~\ref{sec:defects}) can generate such a network pair. By construction, these special architectural transformations do not change the shape of any local loops, and thus do not affect the SS-space (see Sec.~\ref{subsec:response_evolution_sss}). As a consequence, under an externally applied load that is supported by both networks $A$ and $A'$, the stress response of both networks must be identical. Since only the bonds $r$ and $p$ differ between the two networks, the stress spaces of networks $A$ and $A'$ are identical up to the following vectors: the LB-state $\unitvec{\bar{\lbs}}_r^A=\unitvec{r}$, present in network $A$ but not in $A'$, and the LB-state $\unitvec{\bar{\lbs}}_p^{A'}=\unitvec{p}$, present in $A'$ but not in $A$. Since the stress response to external loading that is supported by both networks must be identical, the LB-states $\unitvec{r}$ and $\unitvec{p}$ will therefore not contribute to the network's mutual supported stress responses: the bonds $r$ and $p$ remain unstressed.\\

\section{Mechanical interpretation of evolving LB-states}\label{app:interpretation}

\begin{figure*}
	\centering
	\includegraphics[]{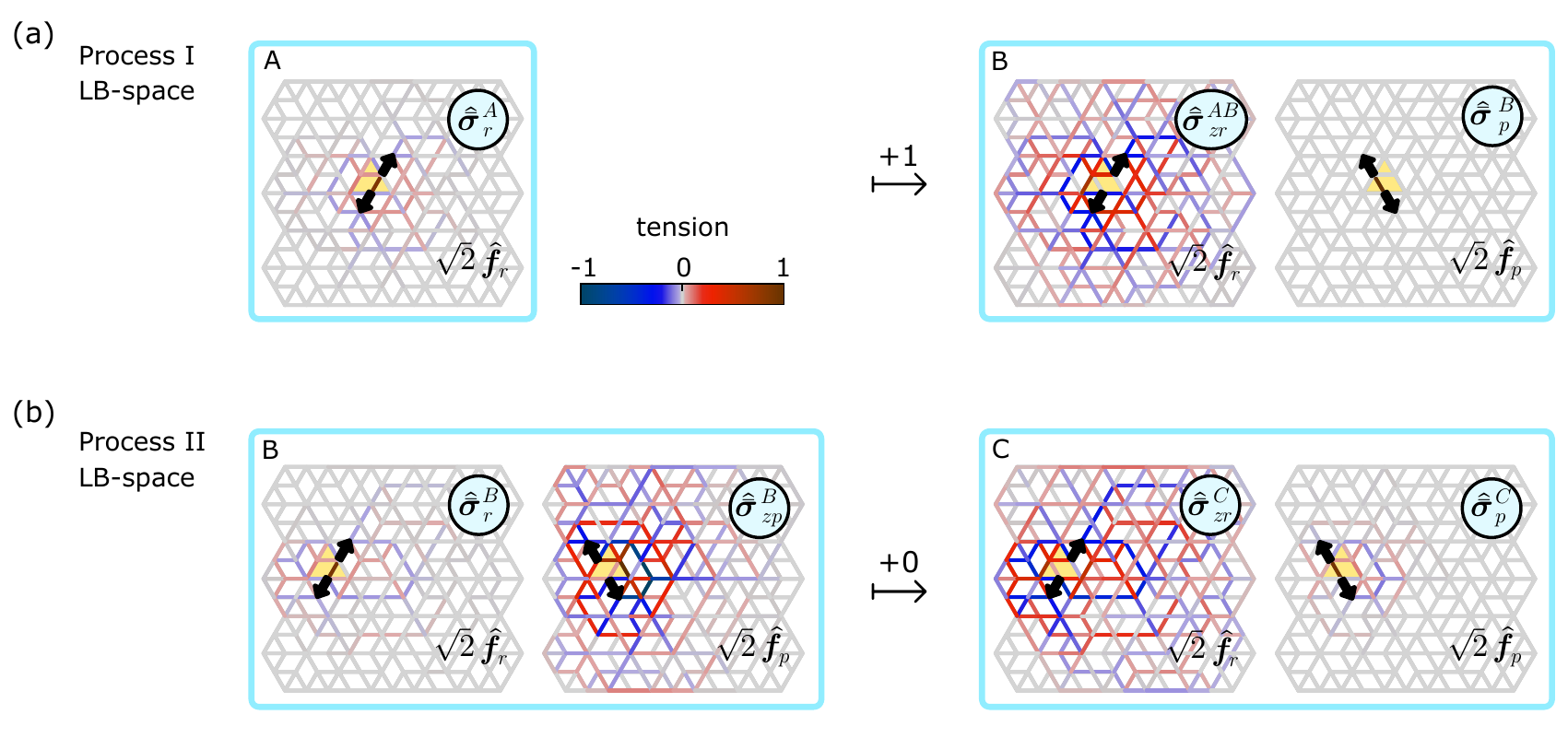}
	\caption{
		Mechanical interpretation of the LB-states that change under a supertriangle rotation for processes I and II.
		(a) A compatible network $A$ transforms into an incompatible network $B$ according to process I. One LB-state $\unitvec{\bar{\lbs}}_r^A$ (colour bar) changes to $\unitvec{\bar{\lbs}}_{zr}^{AB}$ under the transformation; both LB-states map to the same nodal load dipole $\sqrt{2}\unitvec{f}_r$ along bond $r$ (arrows). One LB-state $\unitvec{\bar{\lbs}}_p^B$ is added in network $B$: it maps to the nodal load dipole $\sqrt{2}\unitvec{f}_p$ along bond $p$, which load is not supported in network $A$.
		(b) An incompatible network $B$ transforms into an incompatible network $C$ according to process II. The LB-state $\unitvec{\bar{\lbs}}_r^B$ in network $B$ changes to $\unitvec{\bar{\lbs}}_{zr}^{C}$ in network $C$. Both LB-states map to the same nodal load dipole $\sqrt{2}\unitvec{f}_r$ along bond $r$ (arrows). In addition, the LB-state $\unitvec{\bar{\lbs}}_p^C$ in network $C$ changes to $\unitvec{\bar{\lbs}}_{zp}^{B}$ in network $B$. Both LB-states map to the same nodal load dipole $\sqrt{2}\unitvec{f}_p$ along bond $p$ (arrows).
	}
	\label{fig_SI_interpretation}
\end{figure*}

Having discussed the evolution of LB-space under supertriangle rotations in Appendix~\ref{app:stress_space_evolution}, we now present the mechanical interpretation of the mutated LB-states. We show here that the few stress states that are added, removed, or modified in processes I and II (Secs.~\ref{subsec:response_evolution_process_I} and~\ref{subsec:response_evolution_process_II}) correspond to the metamaterials' stress response to well-defined external nodal loads. In particular, we show below that all mutated LB-states correspond to nodal load dipoles along the two bonds $r$, $p$ that are mutually exclusive between the post- and pre-transformation networks. A nodal load dipole generates equal and opposite forces at two nodes, and is oriented along the connecting line between the two nodes. The mutating LB-states either generate a large stress on a single bond and a diffuse field around it, or an extended stress field around a missing bond, as illustrated in Fig.~\ref{fig_SI_interpretation}.\\

We first consider the stress response evolution of process I, when a compatible material $A$ is transformed into an incompatible material $B$ (Fig.~\ref{fig_SI_interpretation}a). During this transformation, the LB-state $\vec{\bar{\lbs}}_r^A$ of network $A$ changes. The physical interpretation of this stress state is as follows. The state $\vec{\bar{\lbs}}_r^A$ is a linear combination of the SS-state $\unitvec{\bar{\sss}}_r^A$ and the unit bond stress $\unitvec{r}$, such that the final LB-state is orthogonal to the SS-state (Eq.~(\ref{eq:make_lbs})). The unit bond stress corresponds via Hooke's law to a nodal load dipole $\sqrt{2}\unitvec{f}_r$: that is, the two nodes connected by bond $r$ undergo an equal and opposite force, extending the bond (here, the prefactor $\sqrt{2}$ is a consequence of normalization). The SS-state, by definition, generates no nodal loads. Thus, the stress state $\vec{\bar{\lbs}}_r^A$ in network $A$ must map to the nodal load state $\sqrt{2}\unitvec{f}_r$:
\begin{equation}\label{eq:app_lb_sra}
	\vec{\bar{\lbs}}_r^A = \unitvec{r} - (\unitvec{r}\cdot\unitvec{\bar{\sss}}_r^A)\unitvec{\bar{\sss}}_r^A \leftrightarrow \sqrt{2}\unitvec{f}_r
\end{equation}
In network $B$, the LB-state $\vec{\bar{\lbs}}_r^A$ is replaced by a new LB-state $\vec{\bar{\lbs}}_{zr}^{AB}$. It is a linear combination of the SS-state $\unitvec{\bar{\sss}}_r^A$ and the unit bond stress $\unitvec{r}$ such that any stress on $r$ is cancelled out (see Eq.~(\ref{eq:make_lbs_2})). Here, again, the unit bond stress $\unitvec{r}$ maps to the nodal load $\sqrt{2}\unitvec{f}_r$, while the SS-state $\unitvec{\bar{\sss}}_r^A$ generates no load. Hence, in network $B$,
\begin{equation}\label{eq:app_lb_szrab}
	\vec{\bar{\lbs}}_{zr}^{AB} = \unitvec{r} - \frac{1}{\unitvec{r}\cdot\unitvec{\bar{\sss}}_r^A}\unitvec{\bar{\sss}}_r^A \leftrightarrow \sqrt{2}\unitvec{f}_r~.
\end{equation}
Lastly, process I introduces a new LB-state $\unitvec{\bar{\lbs}}_p^B = \unitvec{p}$ in network $B$. Using the same arguments as above, we find that the new LB-state corresponds to a load dipole $\sqrt{2}\unitvec{f}_p$ along bond $p$:
\begin{equation}
	\vec{\bar{\lbs}}_{p}^{B} = \unitvec{p} \leftrightarrow \sqrt{2}\unitvec{f}_p~.
\end{equation}
This LB-state has no counterpart in network $A$: there, the nodal load $\sqrt{2}\unitvec{f}_p$ activates the compatible material's floppy mode, and is not supported.
The remaining LB-states $\{ \unitvec{\bar{\lbs}}_{zr}^A \}$, that are shared between networks $A$ and $B$, are unchanged; they map to identical loads in both networks. An overview of the mutated LB-states, and the nodal loads corresponding to the latter, is shown in Fig.~\ref{fig_SI_interpretation}a.\\

Secondly, we treat the stress response evolution of process II, where an incompatible material $B$ is mutated into an incompatible material $C$ (Fig.~\ref{fig_SI_interpretation}b). There are two LB-states that are modified during this transformation: $\unitvec{\bar{\lbs}}_r^B$ and $\tilde{\unitvec{\bar{\lbs}}}_{zp}^{BC}$ in network $B$ are changed into $\unitvec{\bar{\lbs}}_p^C$ and $\tilde{\unitvec{\bar{\lbs}}}_{zr}^{BC}$ in network $C$. Using an analogous argument as for process I, the LB-state $\unitvec{\bar{\lbs}}_r^B$ in network $B$ maps to the nodal load $\sqrt{2}\unitvec{f}_r$:
\begin{equation}\label{eq:app_lb_srb}
	\vec{\bar{\lbs}}_r^B = \unitvec{r} - (\unitvec{r}\cdot\unitvec{\bar{\sss}}_r^B)\unitvec{\bar{\sss}}_r^B \leftrightarrow \sqrt{2}\unitvec{f}_r
\end{equation}
In intermediate network $BC$:
\begin{equation}\label{eq:app_lb_szrbc}
	\vec{\bar{\lbs}}_{zr}^{BC} = \unitvec{r} - \frac{1}{\unitvec{r}\cdot\unitvec{\bar{\sss}}_r^B}\unitvec{\bar{\sss}}_r^B \leftrightarrow \sqrt{2}\unitvec{f}_r
\end{equation}
and finally in network $C$:
\begin{equation}\label{eq:app_lb_szrc}
	{\vec{\bar{\lbs}}}_{zr}^{C} = \mathrm{Rej}(\vec{\bar{\lbs}}_{zr}^{BC}, \unitvec{\bar{\sss}}_p^C) =
	\unitvec{r}
	- \frac{ \unitvec{\bar{\sss}}_r^B - ( \unitvec{\bar{\sss}}_r^B\cdot \unitvec{\bar{\sss}}_p^C) \unitvec{\bar{\sss}}_p^C
	}{\unitvec{r}\cdot\unitvec{\bar{\sss}}_r^B}
	\leftrightarrow \sqrt{2}\unitvec{f}_r
\end{equation}

Similarly, the LB-state $\unitvec{\bar{\lbs}}_p^C$ maps to the nodal load $\sqrt{2}\unitvec{f}_p$ in network $C$:
\begin{equation}\label{eq:app_lb_spc}
	\vec{\bar{\lbs}}_p^C = \unitvec{p} - (\unitvec{p}\cdot\unitvec{\bar{\sss}}_p^C)\unitvec{\bar{\sss}}_p^C \leftrightarrow \sqrt{2}\unitvec{f}_p
\end{equation}
In intermediate network $BC$:
\begin{equation}\label{eq:app_lb_szpbc}
	\vec{\bar{\lbs}}_{zp}^{BC} = \unitvec{p} - \frac{1}{\unitvec{p}\cdot\unitvec{\bar{\sss}}_p^C}\unitvec{\bar{\sss}}_p^C \leftrightarrow \sqrt{2}\unitvec{f}_p
\end{equation}
and finally in network $B$:
\begin{equation}\label{eq:app_lb_szpb}
	{\vec{\bar{\lbs}}}_{zp}^{B} = \mathrm{Rej}(\vec{\bar{\lbs}}_{zp}^{BC}, \unitvec{\bar{\sss}}_r^B) =
	\unitvec{p}
	- \frac{\unitvec{\bar{\sss}}_p^C - ( \unitvec{\bar{\sss}}_p^C\cdot \unitvec{\bar{\sss}}_r^B) \unitvec{\bar{\sss}}_r^B
	}{\unitvec{p}\cdot\unitvec{\bar{\sss}}_p^C}
	\leftrightarrow \sqrt{2}\unitvec{f}_p
\end{equation}
The remaining LB-states $\{\unitvec{\bar{\lbs}}_{zrzp}^{BC}\}$ are unmodified and map to the same nodal loads in both networks. The mutated LB-states are illustrated in Fig.~\ref{fig_SI_interpretation}b.\\

Lastly, we discuss the stress response evolution for process III, where a compatible material $A$ transforms to a distinct compatible material $A'$%
There are two LB-states that are modified during this transformation: $\unitvec{r}$ and $\unitvec{p}$ are mutually exclusive LB-states of networks $A$ and $A'$ respectively. Using similar arguments as above, the LB-state $\unitvec{r}$ in network $A$ maps to the nodal load dipole $\sqrt{2}\unitvec{f}_r$:
\begin{equation}
	\unitvec{r} \leftrightarrow \sqrt{2}\unitvec{f}_r
\end{equation}
This load dipole is not supported in network $A'$---it activates the global floppy mode of the system---and there is no counterpart to the LB-state $\unitvec{r}$ in network $A'$. Analogously, in network $A'$, 
\begin{equation}
	\unitvec{p} \leftrightarrow \sqrt{2}\unitvec{f}_p~,
\end{equation}
and this LB-state in network $A'$, being unsupported by network $A$, has no counterpart in the LB-space of $A$.

\section{Derivation of the stress response difference}\label{app:stress_response_difference}

With our description of the stress space evolution and its physical interpretation in Appendices~\ref{app:stress_space_evolution} and~\ref{app:interpretation}, we are now in a position to derive exactly how a metamaterial's stress response under external loading changes when its architecture is changed by rotating a supertriangle. In particular, we found that the SS-space of two networks related by a single supertriangle rotation are identical up to at most two mutually exclusive SS-states. Comparing two networks, related by a supertriangle rotation, by calculating their stress response difference $\Delta\vec{\lbs}$ under identical supported loads, we will now show that $\Delta\vec{\lbs}$ is a linear combination of only those SS-states that have been changed by the network's architectural transformation.\\

In any network, the stress response $\vec{\lbs}$ to an arbitrary supported load $\vec{f}$ can be written as a unique linear combination of LB-states: $\vec{\lbs} = \sum_{i=1} \left( C_i \vec{\lbs}_i \right)$, where the set $\{ \vec{\lbs}_i\}$ is any linearly independent basis of stress vectors spanning the LB-space, and the coefficients $C_i$ depend on the applied load, the material's geometry, and the choice of basis. The exact coefficients can be calculated using the matrix formalism discussed in Sec.~\ref{sec:linmech}. We use this representation to find an expression for the stress response difference between two networks, related via process I, II, or III, under identical supported loads.

We first consider networks $A$ and $B$, related via process I. When structure $A$ is subjected to a supported load $\vec{f}$---that is, a load that does not excite the FM of network $A$---the stress response $\vec{\lbs}^A$ is written in a straightforward way:
\begin{equation}\label{eq:app_a} 
	\vec{\lbs}^{A}   = \sum_{i=1}^{N_b-2H-1} \left( C_i \unitvec{\bar{\lbs}}^A_{zr,i} \right) + C_r \vec{\bar{\lbs}}_{r}^{A}~,
\end{equation}
where we have chosen a basis of LB-space such that the LB-states $\{\unitvec{\bar{\lbs}}^A_{zr}\}$ are shared between the two networks, and the LB-state $\vec{\bar{\lbs}}_{r}^{A}$ is unique to network $A$ (see Appendix~\ref{app:stress_space_evolution}). As discussed in Appendix~\ref{app:interpretation}, when a supertriangle is rotated in network $A$ to produce network $B$, the nodal load dipole generated by the stress state $\vec{\bar{\lbs}}_{r}^{A}$ in network $A$ is supported instead by the stress state $\vec{\bar{\lbs}}_{zr}^{AB}$ in network $B$; in addition, the basis of LB-space now contains an extra LB-state $\unitvec{p}$ that maps to a load dipole along bond $p$. For network $B$, the stress response to the same external loading $\vec{f}$ is then written as:
\begin{equation}\label{eq:app_b}
	\vec{\lbs}^{B}   = \sum_{i=1}^{N_b-2H-1} \left( C_i \unitvec{\bar{\lbs}}^A_{zr,i} \right)  + C_r \vec{\bar{\lbs}}_{zr}^{AB}+ C_p\unitvec{p}~.
\end{equation}
Comparing Eqs.~(\ref{eq:app_a}) and~(\ref{eq:app_b}), we note that the LB-states $\{\unitvec{\bar{\lbs}}^A_{zr,i}\}$ are shared between networks $A$ and $B$, and map to identical loads, so that the coefficients $C_i$ are equal. Furthermore, $C_a=0$ by necessity, since the load dipole along bond $p$ excites the FM of network $A$ and cannot be part of our load $\vec{f}$, which must be supported by both networks. Lastly, the stress field $\vec{\bar{\lbs}}_{zr}^{AB}$ corresponds to the stress field $\vec{\bar{\lbs}}_{r}^{A}$---both mapping to the load dipole $\sqrt{2}\unitvec{f}_r$---so that the coefficient $C_r$ in both equations is equal. Using Eqs.~(\ref{eq:app_lb_sra}-\ref{eq:app_lb_szrab}) and Eqs.~(\ref{eq:app_a}-\ref{eq:app_b}), we find the following expression for the stress response difference between networks $A$ and $B$:
\begin{equation}\label{app:eq_stressdiff_ab}
	\Delta \vec{\lbs} = \vec{\lbs}^{B} - \vec{\lbs}^{A}= C_r \frac{-1+(\unitvec{r}\cdot\unitvec{\bar{\sss}}_r^A)^2}{\unitvec{r}\cdot\unitvec{\bar{\sss}}_r^A} \unitvec{\bar{\sss}}_r^A
	\in \mathrm{Sp}(\unitvec{\bar{\sss}}_1^A)~.
\end{equation}
Eq.~(\ref{app:eq_stressdiff_ab}) shows that the stress response difference between the two networks is parallel to the single mutated SS-state $\unitvec{\bar{\sss}}_r^A$. We confirm this finding via numerical calculations: the stress response difference between network $A$ with no defect and network $B$ with a structural defect, illustrated in Fig.~\ref{fig_intro}a (right) corresponds exactly to the lost state of self stress shown in Fig.~\ref{fig_spacechanges}b (top), resulting in a differential stress response that is localized near the defect.\\

A similar procedure allows us to find the stress response difference between two distinct incompatible networks $B$ and $C$, related via process II. The stress response of network $B$ may be written as:
\begin{equation}\label{eq:app_b2}
	\vec{\lbs}^B = \sum_{i=1}^{N_b-2H-1} \left( C_i \unitvec{\bar{\lbs}}_{zpzr,i}^{BC} \right) + C_r \vec{\bar{\lbs}}_{r}^{B} + C_p \vec{\bar{\lbs}}_{zp}^{B}~,
\end{equation}
while the stress response of network $C$ is given by:
\begin{equation}\label{eq:app_c}
	\vec{\lbs}^B = \sum_{i=1}^{N_b-2H-1} \left( C_i \unitvec{\bar{\lbs}}_{zpzr,i}^{BC} \right) + C_r \vec{\bar{\lbs}}_{zr}^{C} + C_p \vec{\bar{\lbs}}_{p}^{C}~.
\end{equation}
Here, the LB-states $\{ \unitvec{\bar{\lbs}}_{zpzr}^{BC} \}$ are shared between networks $B$ and $C$, while the LB-states $\vec{\bar{\lbs}}_{r}^{B}$ and $\vec{\bar{\lbs}}_{zp}^{B}$, that map to load dipoles $\sqrt{2}\unitvec{f}_{r}$ and $\sqrt{2}\unitvec{f}_{p}$ in network $B$, are replaced by their commensurate counterparts $\vec{\bar{\lbs}}_{zr}^{C}$ and $\vec{\bar{\lbs}}_{p}^{C}$ in network $C$, consistent with Appendix~\ref{app:interpretation}. Using Eqs.~(\ref{eq:app_b2}-\ref{eq:app_c}) and Eqs.~(\ref{eq:app_lb_srb}-\ref{eq:app_lb_szpb}), the stress response difference between the two structures then reduces to the following equation:
\begin{equation}\label{app:eq_stressdiff_bc}
	\begin{split}
		\Delta \vec{\lbs} &= \vec{\lbs}^{C} - \vec{\lbs}^{B}\\
		&= \unitvec{\bar{\sss}}_r^B \left[
		C_r \left(\frac{-1+(\unitvec{r}\cdot\unitvec{\bar{\sss}}_r^B)^2}{\unitvec{r}\cdot\unitvec{\bar{\sss}}_r^B}\right)
		+ 
		C_p \left( \frac{-\unitvec{\bar{\sss}}_r^B\cdot\unitvec{\bar{\sss}}_p^C}{\unitvec{p}\cdot\unitvec{\bar{\sss}}_p^C} \right)
		\right]\\
		&+
		\unitvec{\bar{\sss}}_p^C \left[
		C_p
		\left(\frac{1-(\unitvec{p}\cdot\unitvec{\bar{\sss}}_p^C)^2}{\unitvec{p}\cdot\unitvec{\bar{\sss}}_p^C}\right)	+
		C_r
		\left(
		\frac{\unitvec{\bar{\sss}}_r^B\cdot\unitvec{\bar{\sss}}_p^C}{\unitvec{r}\cdot\unitvec{\bar{\sss}}_r^B} 
		\right)	
		\right]\\
		&\in \mathrm{Sp}(\unitvec{\bar{\sss}}_{r}^B,\unitvec{\bar{\sss}}_{p}^C).
	\end{split}
\end{equation}
Once again, the two networks' stress response difference is contained in the space spanned by their two mutually exclusive SS-states, $\unitvec{\bar{\sss}}_{r}^B$ and $\unitvec{\bar{\sss}}_{p}^C$. Note that the stress response difference of Eq.~(\ref{app:eq_stressdiff_ab}) (process I) is a special case of the general expression in Eq.~(\ref{app:eq_stressdiff_bc}) for process II.

Consider finally the two compatible networks $A$ and $A'$, related via process III. With the same procedure as for processes I and II, we can write:
\begin{equation}\label{eq:app_az}
	\vec{\lbs}^A = \sum_{i=1}^{N_b-2H-1} \left( C_i \unitvec{\bar{\lbs}}_{zr,i}^{A} \right) + C_r \unitvec{r}~,
\end{equation}
while the stress response of network $C$ is given by:
\begin{equation}\label{eq:app_azprime}
	\vec{\lbs}^{A'} = \sum_{i=1}^{N_b-2H-1} \left( C_i \unitvec{\bar{\lbs}}_{zr,i}^{A} \right) + C_p\unitvec{p}
\end{equation}
By definition, under a load that is supported in both networks, the coefficients $C_r$ and $C_p$ must be zero (see Appendix~\ref{app:interpretation}); and hence, there is no stress response difference between the two structures $A$ and $A'$ under identical, supported loads. Again, the stress response difference for process III is a special case of Eq.~(\ref{app:eq_stressdiff_bc}) for process II.

In conclusion: the stress response difference between two networks (related by a single supertriangle rotation) under identical, supported loading  is contained in the span of the structures' mutually exclusive SS-states. There may be zero, one, or two such states, corresponding to processes III, I, and II respectively. The precise magnitude of the stress response difference can be found using Eqs.~(\ref{app:eq_stressdiff_ab}) (process I) and~(\ref{app:eq_stressdiff_bc}) (process II); the stress response difference for process III is trivially zero.

\end{document}